\documentclass[onecolumn]{aastex61}
\pdfoutput=1

\usepackage{array}
\usepackage{needspace}

\newcolumntype{P}[1]{>{\raggedright\let\newline\\\arraybackslash\hspace{0pt}}m{#1}}
\newcommand{\mydeg}{{$^{\circ}$}}

\newcommand{\etal}{{\it et~al.}}

\newcommand{\musn}{Arrokoth} 

\tosubmitjournal{Astronomical Journal, Version 1.1, 2019/12/30}
\shortauthors{Buie et al.}
\shorttitle{Arrokoth Occultations}

\begin{document}

\title{Size and Shape Constraints of (486958) Arrokoth from Stellar Occultations}

\correspondingauthor{Marc Buie}
\email{buie@boulder.swri.edu}

\AuthorCollaborationLimit=2

%
%
%
%
% MACHINE GENERATED FILE, DO NOT EDIT!!!!
%
%
%
%
%
%
% MACHINE GENERATED FILE, DO NOT EDIT!!!!
%
%
%
%
%
%
% MACHINE GENERATED FILE, DO NOT EDIT!!!!
%
%
%
%
%
% built from author_20191221.tsv and authorlist.tex
% generated at Sat Dec 21 15:18:13 2019 UTC
%
\author[0000-0003-0854-745X]{Marc W. Buie}
\affiliation{Southwest Research Institute\\
1050 Walnut St., Suite 300, Boulder, CO 80302 USA}

\author[0000-0003-0333-6055]{Simon B. Porter}
\affiliation{Southwest Research Institute\\
1050 Walnut St., Suite 300, Boulder, CO 80302 USA}

\author{Peter  Tamblyn}
\affiliation{Southwest Research Institute\\
1050 Walnut St., Suite 300, Boulder, CO 80302 USA}

\author[0000-0001-8406-4172]{Dirk  Terrell}
\affiliation{Southwest Research Institute\\
1050 Walnut St., Suite 300, Boulder, CO 80302 USA}

\author[0000-0002-6722-0994]{Alex Harrison Parker}
\affiliation{Southwest Research Institute\\
1050 Walnut St., Suite 300, Boulder, CO 80302 USA}

\author[0000-0002-1785-5262]{David  Baratoux}
\affiliation{G\'eosciences Environnement Toulouse, UMR5563 CNRS, IRD and University of Toulouse\\
14 Avenue Edouard Belin, 31400, Toulouse, France}

\author{Maram  Kaire}
\affiliation{MESRI\\
All\'ees Papa Gueye Fall Dakar, S\'en\'egal}
\affiliation{ASPA\\
15, nord Foire Dakar, S\'en\'egal}

\author[0000-0002-6477-1360]{Rodrigo  Leiva}
\affiliation{Southwest Research Institute\\
1050 Walnut St., Suite 300, Boulder, CO 80302 USA}

\author[0000-0002-3323-9304]{Anne J. Verbiscer}
\affiliation{University of Virginia, Department of Astronomy\\
P. O. Box 400325,  Charlottesville, VA 22904 USA}

\author{Amanda M. Zangari}
\affiliation{Southwest Research Institute\\
1050 Walnut St., Suite 300, Boulder, CO 80302 USA}

% ...

\begin{abstract}

We present the results from four stellar occultations by (486958)
Arrokoth, the flyby target of the New Horizons extended mission.
Three of the four efforts led to positive detections of the body, and
all constrained the presence of rings and other debris, finding none.
Twenty-five mobile stations were deployed for 2017 June 3 and augmented
by fixed telescopes.  There were no positive detections from this effort.
The event on 2017 July 10 was observed by SOFIA with one very short chord.
Twenty-four deployed stations on 2017 July 17 resulted in five chords
that clearly showed a complicated shape consistent with a contact binary
with rough dimensions of 20 by 30 km for the overall outline.  A visible
albedo of 10\% was derived from these data.  Twenty-two systems were
deployed for the fourth event on 2018 Aug 4 and resulted in two chords.
The combination of the occultation data and the flyby results provides
a significant refinement of the rotation period, now estimated to
be 15.9380 $\pm$ 0.0005 hours.  The occultation data also provided
high-precision astrometric constraints on the position of the object that
were crucial for supporting the navigation for the New Horizons flyby.
This work demonstrates an effective method for obtaining detailed size
and shape information and probing for rings and dust on distant Kuiper
Belt objects as well as being an important source of positional data
that can aid in spacecraft navigation that is particularly useful for
small and distant bodies.

\end{abstract}

\section{Introduction}

The New Horizons extended mission target was (486958) Arrokoth,
previously known as 2014~MU$_{69}$ \citep{ste18}.  This cold classical
Kuiper Belt object was discovered in 2014 by a targeted search
\citep{bui18,por18} with the Hubble Space Telescope (HST).  Through the end of
2018 we continued to follow \musn\ to collect astrometry and photometry
with HST\null.  The mean apparent magnitude is $R=27$ mag with an absolute
magnitude of $H_R=11$ \citep{bui18,ben19}.
It is important to realize \musn\ is at the limit of
capability of HST and no ground-based facility has successfully detected \musn.
In the case of HST, the detection limit is set by the size of the telescope.
From the ground, larger telescopes are available but the increase in light
gathering power is lost due to the poorer image quality
imposed by the atmosphere combined with the extremely crowded background
stellar field.  There are very effective techniques for removing the stellar
background but these techniques cannot remove the noise introduced by the stars.
This noise component effectively dictates the limiting magnitude of the
subtracted images.  Data with better seeing do reach fainter limits but
our best ground-based search data missed detecting \musn\ by about one
stellar magnitude.

The New Horizons mission team needed as much information about the target
as possible prior to the flyby of \musn\ on 2019 Jan 01 (UT).  The spatial
resolution of HST is about 1200~km per pixel in our imaging data.
Stellar occultations provide higher spatial resolution data, typically at or
better than 1~km, that greatly exceed what is possible with HST\null.
However, a successful occultation has its own challenges.  The target body
must pass close enough to a star when it can be seen from somewhere on
Earth that has good weather.  We also must be able to accurately predict
where the shadow will be so that telescopes can be deployed to the correct
location to record the event.  With an object as small as \musn\ the
probability of its shadow crossing a fixed observatory is very, very low.
Given a suitable opportunity, an occultation can do two important things
for a mission.  First, we are able to measure the projected area of the
body and thus infer its albedo when combined with its absolute magnitude,
provided enough suitably placed stations can observe the event.
Measuring the albedo was important
to New Horizons as input to the design of the imaging sequences to know
the signal-to-noise ratio that a given observation would yield.  Second,
the occultation data provide astrometry.  At the time of the occultation,
we have precise knowledge of the position of the body relative to the star.
With a sufficiently accurate star position, this information translates to
astrometric data that is at least a factor of 100 better than a single
HST image and subject to completely different potential systematic errors.
These astrometric constraints were expected to be very important for improved
orbit estimates prior to the encounter for both navigation of New Horizons
as well as pointing information for the cameras.

The albedo of \musn\ was clearly one of our measurement objectives but it
also played a role in building a successful observing strategy for the
deployment for the occultation observations.  Using our absolute magnitude
estimate of $H_V=11.1$, a 4\% albedo implies a diameter of 40~km.  Ignoring the
photometric errors on the absolute magnitude, this represented a practical
upper limit to its size.  A lower limit on size was harder to pin down,
but at 20\% albedo, the diameter would have been 20~km.
This plausible range in size combined with the uncertainty of the
orbit estimation played a strong role in
the occultation deployment.  The heliocentric distance of \musn\ in
2017 was 43.3 AU.  At that distance, the scale on the plane of the sky
was 31.4~km/mas.  These spatial scales required knowing the position
of the object and the star to at least 1~mas for a
reasonable chance at a successful multi-chord occultation.

The orbit for \musn\ indicates that it is a cold-classical Kuiper Belt object
\citep[$a=44.4$, $e=0.038$, $i=2.45$;][]{por18}.  Observations of other
cold-classicals reveal a population with a very high fraction of equal-mass binary
objects \citep{nol08,nes10,fra17}.  The HST observations showed no signs of
binarity, but we expected an occultation to probe at a much smaller spatial scale.

We present here a description of results for three stellar occultation observation
campaigns in 2017 and one campaign in 2018.  All of the campaigns returned useful
data to constrain the size, shape, and orbit of the New Horizons extended mission
target.  

\section{General Background on Events}

Our first step for this project was to search the USNO CCD Astrograph Catalog
(UCAC4) \citep{zac13} for candidate stars for occultations in 2017.  This
search provided a list of three good candidate stars.  The positional
uncertainties on these stars were too high to get a useful prediction, but
they did support the earliest stages of planning.  Based on this initial
information, we requested time on NASA/DLR's Stratospheric Observatory
for Infrared Astronomy (SOFIA), which the telescope allocation committee
approved.  However, further analysis indicated that SOFIA could only support
one of the three opportunities because of logistical constraints.  Also at
this time, we applied for time on other large telescope facilities and put
the word out to the community about these opportunities.

Other than SOFIA, which is mobile, the large telescope facilities were unlikely
to be in the right place for a solid-body event.  However, all of them could be
useful for probing the \musn\ Hill sphere for additional material, especially
rings or extended dust structures.  No such material was found, and those
results are described further in \citet{you18}, which excluded rings with radii
up to 1000 km and widths of greater than 720 m.  The key to success
for a solid-body detection was a large number of mobile ground stations.

Based on our estimates of the final prediction uncertainties we built
a plan for 25 mobile stations.  New Horizons procured 22 systems
for this project that are based at Southwest Research Institute (SwRI) in
Boulder, Colorado.  These identical systems were each assigned a system
code between T01 and T22.  In addition, the University of Virginia (UVA) provided
three additional systems, all of differing designs.  We assigned these systems
codes T23 to T25 and we will describe these separately.

We relied heavily on astrometric support catalogs for this project.
During the project we used many different catalogs.  However, getting a good
prediction required using the same catalog to calibrate the HST astrometry
and obtain the position of the occultation star.
In the earliest phase of this project, we used a special
catalog developed by S. Gwyn at Canadian Astronomy Data Centre (CADC) using
data from the Canada France Hawaii Telescope (CFHT) Megacam
system.  This catalog had better internal consistency than any other
catalog available at the time; however, this catalog did not have
useful proper motion information. To overcome this limitation, we used
mean apex proper motion corrections \citep{gwy14}. This mean correction 
was acceptable for the orbit estimation but was inadequate for the
positions of the occultation stars themselves. 

The release of the Gaia Data Release 1 (DR1) catalog \citep{gai16} allowed us
to revise and improve the support catalog and positions of the occultation stars.
Essentially all of the stars we used from the catalog were too faint to
have proper motions from DR1, so we were again forced to use mean proper
motions.  We made an additional effort to search for other epochs of data on
these fields to constrain the proper motions of the occultation stars.  We
found data in the MACHO (MAssive Compact Halo Object) archives \citep{all01}
which provided a slightly improved set of predictions; however, those
predictions were still inadequate.

We obtained HST images of all candidate occultation stars.  The HST images
showed no signs of stellar duplicity.  The images captured the positions of
the stars near the epoch of the occultation so that the projected uncertainty
from proper motion would not dominate the prediction uncertainty.  As we were
working to extract this information, the Gaia Mission graciously agreed to
provide pre-release positions from the Gaia Data Release 2 (DR2) catalog which
had just finished its initial processing.  With DR2 in hand, there was no need
to get positions from the HST data.

The Gaia DR2 pre-release sub-catalog covered an area of the sky encompassing
all of the HST observations of \musn\ from the discovery epoch in 2014 through
the end of 2017.  The area also included the occultation stars.  This
catalog contained proper motions and uncertainties for all listed stars.
The catalog density in these areas was high enough that the final uncertainty
of the WCS calibration
for the HST images was a negligible component of the occultation predictions.
Having Gaia DR2 information on the occultation stars was
fundamentally important because it meant all of the astrometry
for \musn\ and all of the occultation stars were in the same
catalog system.  More significantly, DR2 was referenced to the same ICRF
as used for navigation of New Horizons.
This allowed us to quantify the uncertainties for the HST observations,
the orbit estimation,
and for the occultation predictions themselves, and ultimately provide useful
positional data that would aid with navigation for the New Horizons flyby.
Table~\ref{tbl-stars} provides the final positions of the stars used
in support of the occultation campaigns.  The first line for each star
provides the full DR2 catalog entry (epoch=2015.5).
The second line contains the positions at
the epoch of the event as well as the propagated uncertainties.
For the catalog positions, we tabulate the parallax, proper motions in
right ascension (PM$_{\alpha}$) and declination (PM$_{\delta}$), and
the Gaia ``G'' magnitude.  More information can be found about the Gaia
catalog values in \citet{gai18}.
Although the uncertainties of all stars were
much better compared to any prior occultation event work, 
these low uncertainties at the epoch of the events were still important.
We will return to this point later in Section~\ref{sec-combine}.
In particular, the star from 2018 is clearly much closer than the rest
as seen by its higher parallax and proper motion.  Without the Gaia results,
this last star would have been completely impossible given the tight targeting
requirements for these occultation attempts.

\begin{deluxetable}{cDccccrrrc}
%\footnotesize
\tablecaption{Occultation Star Data\label{tbl-stars}}
\tablewidth{0pt}
\tablehead{
\colhead{Star}&
\multicolumn2c{Epoch}&
\colhead{R.A.($\alpha$)}&
\colhead{$\sigma_{\alpha}$}&
\colhead{Dec.($\delta$)}&
\colhead{$\sigma_{\delta}$}&
\colhead{Parallax}&
\colhead{PM$_{\alpha}$}&
\colhead{PM$_{\delta}$}&
\colhead{$G$}\\
&
\multicolumn2c{(year)}&
\colhead{(deg)}&
\colhead{(mas)}&
\colhead{(deg)}&
\colhead{(mas)}&
\colhead{(mas)}&
\colhead{(mas/yr)}&
\colhead{(mas/yr)}&
\colhead{(mag)}
}
%\colnumbers
\decimals
\startdata
MU20170603& 2015.5   & 285.8937182917& 0.040& $-$20.5775960556& 0.039& 0.326$\pm$0.053& $-$0.504$\pm$0.078&    0.567$\pm$0.071& 15.27\\
{        }& 2017.419 & 285.8937182238& 0.158& $-$20.5775957188& 0.142& \multicolumn{1}{c}{\nodata}& \multicolumn{1}{c}{\nodata}& \multicolumn{1}{c}{\nodata}& \nodata\\
\hline
MU20170710& 2015.5   & 285.1734150250& 0.040& $-$20.6457042778& 0.038& 0.493$\pm$0.042&    2.988$\pm$0.081& $-$0.922$\pm$0.075& 15.53\\
{        }& 2017.520 & 285.1734168060& 0.169& $-$20.6457047916& 0.156& \multicolumn{1}{c}{\nodata}& \multicolumn{1}{c}{\nodata}& \multicolumn{1}{c}{\nodata}& \nodata\\
\hline
MU20170717& 2015.5   & 285.0345477417& 0.048& $-$20.6605479583& 0.047& 0.506$\pm$0.056&    0.451$\pm$0.090& $-$4.696$\pm$0.078& 12.75\\
{        }& 2017.539 & 285.0345479854& 0.190& $-$20.6605506170& 0.166& \multicolumn{1}{c}{\nodata}& \multicolumn{1}{c}{\nodata}& \multicolumn{1}{c}{\nodata}& \nodata\\
\hline
MU20180804& 2015.5   & 286.0894861917& 0.023& $-$20.5934747944& 0.022& 2.486$\pm$0.026&    8.669$\pm$0.046& $-$12.310$\pm$0.043& 13.381\\
{        }& 2018.589 & 286.0894938000& 0.144& $-$20.5934853695& 0.135& \multicolumn{1}{c}{\nodata}& \multicolumn{1}{c}{\nodata}& \multicolumn{1}{c}{\nodata}& \nodata\\
\enddata
\tablecomments{\scriptsize
Positions are all referenced to EME2000.
}
\end{deluxetable}

Another important component of event predictions is the orbit estimation
for \musn.  We had an on-going program with HST to observe the object
periodically and collect additional astrometry, starting in 2014 with its
discovery and continuing through October 2018.  Our baseline observing cadence
was five epochs of data per year, spread out over the apparition.  Each epoch
consisted of five 370-sec exposures, usually within a single visibility
window from HST (one orbit).  In 2017, a special lightcurve
campaign added an additional 24 orbits of astrometric data just prior
to the SOFIA occultation attempt.

\section{Mobile Instrumentation}

\subsection{2017 Summary}

\subsubsection{SwRI-NH Systems, T01-T22}

We assembled the twenty-two mobile T01-T22 systems with commercially available
components plus custom storage and shipping crates.
Each system included a Skywatcher 16'' (40-cm) Dobsonian telescope with
computerized drive electronics.  Each telescope provided an f/4.4 beam at the
Newtonian focus.  Each telescope's secondary housing is in a short tube held up from
the primary support tube by three rods.  The secondary housing collapses down next
to the primary tube for storage and transportation.  The alt-az drive system design
allowed us to move the telescope either manually or with the motors without loss of pointing.
Once properly aligned, the telescope automatically tracks a point on the sky but the
field rotates slowly on the detector as it tracks.
Because these telescope systems did not have GPS installed, each system required manual entry
of time and position at the start of each observing session.  The optics were reasonably
robust but did require some attention and re-collimation
with each use.  The primary mirror support rarely needed realignment but we found
that the secondary inevitably moved during transport.
A laser collimator became an essential component of each field support kit.  We also learned
that these telescopes are susceptible to stray light interfering with the camera.
In addition, relatively light winds can shake the telescope: image motion becomes apparent
at 5 mph (8 kph); we found it very difficult to use the telescope at all over 10 mph (16 kph)
without a mitigation strategy.

We chose a QHY174M-GPS camera, using a thermo-electrically cooled CMOS detector
with a built-in GPS receiver.  The camera's array size is 1920x1200 pixels and
provides a field of view of 21x13 arcmin with a pixel size of 0.67 arcsec.
We used SharpCap\footnote{\url{https://www.sharpcap.co.uk}} to readout the detector and save the data.
This software can write each image to a separate FITS file while also
recording the GPS-based start time for each exposure.  The software writes the latitude
and longitude to the header of each file.  Because the camera does not pass out the altitude,
we had to manually record the altitude information.  All of our systems used
SharpCap version 3.0.3938.0 for the entirety of the 2017 campaigns.  The GPS position in this version
of the software was not always accurate, so we instructed all teams 
to use another means (usually a cell phone) to record their location.
The camera has a fast 12-bit A/D converter but SharpCap shifts the data to
the most significant bits of a 16-bit integer.  This version of SharpCap 
allowed a variable system gain setting between 1 and 48; the higher the gain number,
the fewer photons per count.  As the gain setting is increased,
the read-noise of the detector decreases while also reducing the dynamic range.
The highest effective gain setting was 30, above which the only
meaningful change is to further reduce the dynamic range.
An approximate gain was 0.004 electrons/DN at GAIN=30.  The read-noise was around 2-3 electrons. We used a temperature set point of 0\mydeg C. With this setting, the dark current is negligible.

We verified the timing information
of the system by analysis using the SEXTA (Southern EXposure Timing Array) system \citep{bar15}.
As long as there is a valid GPS fix and an up-to-date leap-second almanac, we found
the QHY camera always had the correct start time within the 2 ms precision of the SEXTA.
However, the camera captures very little supporting information with the data,
and in particular it does not always capture the state of the GPS and almanac
download information.  Updated firmware was eventually provided but too late to
be used for these occultation campaigns.  These cameras also do not appear to save
the almanac information between uses; therefore, we ran the cameras for
at least 20 minutes so that we were certain to get the correct leap second information.
Before the update, the system time was off by 2 seconds.
There were no direct indications that this update happened without
watching the clock very carefully.  The observing protocol for these
systems naturally led to a period of operation before data collection well
in excess of the almanac update interval.

We collected data with an inexpensive laptop with a spinning hard
disk.  This system was not quite capable of 5 Hz read-out speeds for full frames, but laptops with solid-state hard drives showed much higher readout speeds, closer
to 12 Hz.  To enable a faster readout speed, we reduced the number of rows read
and saved. This also reduced the field of view.  Changing the number of columns
made little difference to the speed.

We ran the laptop from its built-in battery in the field.  We powered the telescope
and camera cooler from a re-chargeable sealed lead-acid battery pack.  Our observing sessions
were rarely longer than 4 hours and the battery capacities were more than adequate for this usage.

We had some variability in overall system performance including a few failures in the field.
During the initial transport, the T22 system mirror detached from its steel support structure.
We used this system as a source of spare parts during the rest of the 2017 deployment.   A few of the
telescope systems had significant amounts of backlash in the gears.  With care and the
calibrations and adjustments noted above, these systems worked sufficiently well for our needs.

\subsubsection{University of Virginia Systems, T23-T25}

The University of Virginia supplemented the 22 SwRI systems with three
additional telescopes: a 24" Dobsonian \textit{f}/4.2 telescope from
Hubble Optics (T23); a 14" Meade LX200-GPS fork-mounted telescope (T24);
and a 14" Celestron EdgeHD telescope mounted on a CGE-Pro equatorial mount (T25).
The two 14" telescopes employed the same QHY174M-GPS sCMOS camera described
previously while the 24" telescope used a higher performance PCO Gold 4.2
sCMOS camera.

T23: The 24" Hubble Optics telescope had more than twice the collecting
area of the next largest telescopes in the network.  A PCO Gold 4.2 2048x2048
sCMOS camera further augmented the sensitivity of this larger aperture by
providing $<$1 e$^-$ read noise exposures at high frame rate.  The 6.5~$\mu$m pixels
of the PCO camera provided a pixel scale of 0.6 arcsec/pixel and a field
of view of 20.5$\times$20.5 arcminutes. Binning of the PCO images produced 1024$\times$1024 
frames with 1.2 arcsecond pixels.

T24: An f/6.3 Meade focal reducer provided a pixel scale of 0.55$''$/pixel
for a QHY camera at the focal plane of the 14" Meade telescope.
The resulting field of view was 17.6$\times$11 arcminutes. 

T25: A Starizona Hyperstar on the 14" Celestron EdgeHD telescope provided an 
\textit{f}/1.9 prime focus that delivered a pixel scale of 1.77$''$/pixel on the QHY array.
The full field of view of the QHY camera in this configuration
was 57$\times$35 arcminutes.

\subsection{2018 Summary}

All of the systems once again used the QHY cameras with embedded GPS receivers.
SharpCap was again used for data collection after upgrading to version 3.1.5219.0.
With this version, the error in recording the position was fixed.  Most
operations were the same as before but the gain control values were a factor
of 10 higher in the new software.  Thus, our previous ``standard'' gain value of
30 was now 300.  Additionally, the GPS receiver status was now visible all the
time on the main control screen to help monitor its state more closely.

\needspace{1\baselineskip}
\subsubsection{SwRI-NH Systems, T01-T22}

We used the same systems for both the 2017 and 2018 events after minor repairs
such as replacing the mirror for T22. We shipped the T01-T19
systems by ocean freight to S\'en\'egal and the T20-T22 systems via air freight
to Bogot\'a, Colombia.  Two telescopes failed in the field because they were
unable to point and track under computer control.  However, we still used
these systems by manually pointing the
telescope to the proper altitude and azimuth so that we had the occultation star in
the field of view at the time of the occultation.

\subsubsection{University of Virginia Systems, T23-24}

University of Virginia's systems for this deployment consisted of two identical
Celestron Edge HD 14" telescopes on CGX-L equatorial mounts.  The optical
properties of these systems, which included Starizona Hyperstar prime-focus
adapters, were identical to the Celestron 14" system (T25) used in the 2017
events.  This once again provided a scale of 1.77$''$/pixel on the 5.86~$\mu$m pixels
of the QHY174M-GPS sensor.  Note that the system IDs for the UVA equipment
from 2018 do not match the system IDs from 2017.

\section{2017 June 3 Event}

This event was the hardest because it was both the first attempt with the new field
systems and because we had relatively poor orbit constraints on \musn.  The initial
rough prediction indicated that we could observe the event from both South America
and southern Africa.  The predicted location shifted significantly in the months
leading up to the event.  However, from the beginning our overall plan involved
splitting our resources between the two continents to improve our chances of 
getting useful data.  Figure~\ref{fig-0603globe} shows the global view of the final
prediction.  We limited the deployment to ground stations because the expected
uncertainty was too high for consideration for a SOFIA flight. 

\begin{center}
\includegraphics[scale=1.0]{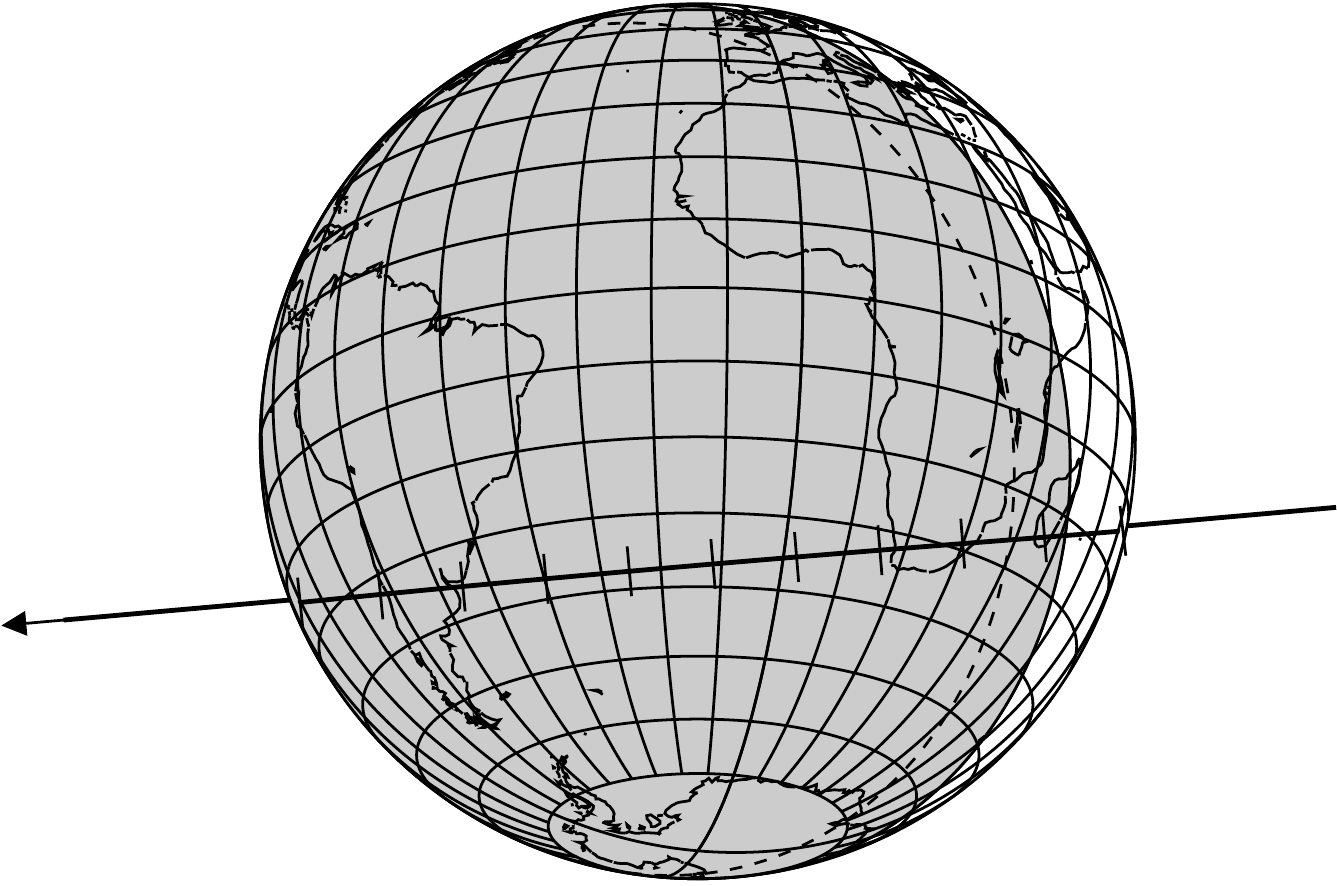}
\figcaption{\label{fig-0603globe}
Global view of 2017-06-03 occultation prediction.
The figure shows the Earth as seen from \musn\ at the time of
geocentric closest approach.  The Sun is below the horizon in the
regions shaded gray.  The dashed line indicates -12$^\circ$ Sun altitude.
The solid line indicates the predicted ground-track with
the width drawn to scale for a 30 km diameter object.  The arrowhead indicates the direction of motion and the ticks are spaced at
one minute intervals from 03:07 to 03:17 UT.  Shadow velocity was
20.0 km s$^{-1}$.  A 66\% illuminated Moon was 103\mydeg\ away from the target
at the time of the event but was below the horizon in Africa.
}
\end{center}

\subsection{Prediction}

The prediction for this event was finalized very close to deployment.
We should have had new astrometry from HST in March 2017 but those
observations were lost to an HST schedule interruption due to an unrelated
technical anomaly.  The earliest we were able to
reschedule HST was 2017 May 1.  Until we obtained the 2017 May 1 data,
the most recent observation data was from 2016 Oct 24. 
The new data provided a substantial increase in the total arc-length
of the \musn\ astrometry, from 2.3 years to 2.9 years with
a corresponding decrease in the extrapolation to the time of the event
from 7 months down to just one month.  Also on May 1, we obtained the HST
observations of the three 2017 target stars.  We used these target star
observations for astrometry and to search for stellar duplicity.  We did
not find any stellar companions or duplicity in the HST images down to the
resolution limit of the data ($\sim$40 mas).

We were able to significantly improve the orbit
estimate with this new astrometry; however, we still needed to resolve fundamental
questions about the position of the
occultation star and the associated uncertainty.
We began working on the HST data to improve our constraints on
the star positions. However, on 2017 May 6 we were provided access to
preliminary data from the planned Gaia DR2 \citep{gai16}.  Those
data included the occultation stars among the stars in the region around \musn\ back to 2014.
We reprocessed all of the HST data with updated reference stars to
improve the orbit estimate.
Within a week we had refined the event prediction enough so that we could determine
where to deploy the observing teams and begin the process of shipping
equipment and setting up travel logistics.  We obtained one more epoch of data from
HST on 2017 May 25, just a day before the first teams left.
We completed the final prediction a couple of days before the June 3 event
with a cross-track uncertainty of 44 km. The in-track (timing) error was 67~km
(3.3 sec).  Unless otherwise stated, all uncertainties stated in this work are 1-$\sigma$ values.

\subsection{Deployment}

We had 24 mobile stations available for deployment.  The equipment was all
sent via air freight to Argentina and South Africa due to the extremely
tight schedule.  Local movement of the systems was handled by individual
vehicles carrying one team and system.

Even with this large number of stations, we could not cover all possible
cases for the object \textit{e.g.} small versus big given the prediction
uncertainty.  To guide the deployment process we used a Monte Carlo simulation.
The simulation uses the cross-track positions for the observing locations relative
to the prediction.  The model employs a circular representation of the occulting
body with an adjustable diameter.  For a given size and set of observing locations,
we draw a random location for the centerline from a normal distribution
consistent with the prediction and its uncertainty.  For a given draw, we compute
the chord length for each site (or note a miss) and record the number
of chords seen.  To be counted, we required a chord to be no shorter
than half the diameter.  This adjustment recognized that we might not see
very short grazing chords given the anticipated noise in the data.
The tool also provides additional provision for a small random component to the
site location.  We could always indicate a desired location to a team, but local
constraints could force them to set up some distance away from the desired location.  By using this
extra random component, we were able to give guidance on how close each team
needed to be to their assigned location.  For this event, the teams needed to
observe from within a 1~km region centered on the assigned location.
After running 10,000 trials, we then generate a histogram as a function of the
number of chords from which to evaluate a given scenario.

A baseline goal for this event was to observe or rule out the largest size
based on a 4\% albedo.  The deployment strategy was guided by the desire to
obtain a strong constraint using just one set of stations (either T01-T12 or
T13-T25 but not both).  A spacing of 15.5 km between sites was chosen so that we
would have no more than a 3\% chance of a null result (zero chords).  This
spacing covered
a range of $\pm1.9\sigma$ or $\pm83$~km and had a 93\% chance of getting two
or more chords.  Given the 44~km cross-track uncertainty for this event, it
would have taken a much larger number of mobile stations to address a smaller
object scenario and we had to accept a poorer constraint for that case.  With
a half-space shift between Argentina and South Africa, the net spacing if all
sites participated in the optimum plan would have been 7.8 km.  The same
pattern would have had a 5\% chance of a null result but an 84\% chance of a
single-chord outcome on a 20~km object.  This tool was very effective in
guiding the Mendoza, Argentina area deployment where the teams had a great
deal of flexibility without the need for detailed site selection scouting
days before the event.  We made adjustments up to the last hours before teams
left for their sites.  The teams in South Africa required more advance warning
due to more complex logistics for site access.  We were able to use this same
tool to assess the outcome for the actual site locations after the event.

\begin{deluxetable}{cP{5 cm}cccccl}
%\footnotesize
\tablecaption{Mobile Observing Stations and Teams for 2017 Jun 03\label{tbl-jun3sta}}
\tablewidth{0pt}
\tablehead{
\colhead{ID}&
\colhead{Team}&
\colhead{Latitude}&
\colhead{E Longitude}&
\colhead{Elevation}&
\colhead{FWHM}&
\colhead{Sky}&
\colhead{Comments}\\
& &
\colhead{(deg)}&
\colhead{(deg)}&
\colhead{(m)}&
\colhead{(pixels)}&
\colhead{(counts)}
}
%\colnumbers
\decimals
\startdata
T01& M. Buie, A. Ocampo, S. Makarchuk&                               $-$33.046832& $-$68.325955&  627&  7.4&  5411& poor tracking at mid-time\\
T02& J. M. Pasachoff, M. Lu, J. Jewell, S. Gurovich&      $-$33.609167& $-$69.006944&  882&  7.3&  6460& \\
T03& C. Olkin, R. Reaves&                              $-$33.946734& $-$67.981331&  604&  8.6&  4041& flares from traffic\\
T04& W. Hanna, C. Erickson, A. Soto&                   $-$33.053972& $-$68.778343&  826&  7.9& 12097& \\
T05& A. Parker, K. Getrost&                            $-$33.649326& $-$68.058762&  582&  9.6&  4214& \\
T06& J. Dunham, P. Tamblyn&                            $-$33.218226& $-$68.612913&  720&  9.4&  6999& \\
T07& D. Dunham, A. Olsen&                              $-$34.011530& $-$69.089410& 1215& 10.2&  5033& \\
T08& S. Slivan, R. Venable&                            $-$32.747256& $-$68.479500&  598&  6.2&  6038& \\
T09& D. Duncan, A. Friedli&                            $-$32.851725& $-$68.392300&  640&  6.9&  8026& flares from traffic\\
T10& S. Conard, B. Keeney, J. Rabassa&                  $-$33.309463& $-$68.900784&  938&  7.1&  5375& \\
T11& L. Wasserman, S. Moss, M. Camino&                 $-$32.564267& $-$68.672067&  600&  8.0&  4549& \\
T12& S. Levine, C. Zuluaga&                            $-$34.100796& $-$67.942469&  559&  6.8&  5612& \\
T13& S. Porter, C. Danforth &                           $-$32.001628& $+$18.777307&   91&  6.2&  3054& some clouds\\
T14& A. Zangari, C. Carter&                            $-$31.524233& $+$23.589731& 1346&  6.1&  3108& \\
T15& C. Tsang, R. Smith&                               $-$31.501944& $+$18.912778&  246&  6.1&  3821& \\
T16& E. Young, A. Rolfsmeier&                          $-$32.352265& $+$18.937847&  146&  5.1&  1707& \\
T17& J. Regester, E. Kramer&                           $-$32.121767& $+$19.054971&  496&  4.1&  1372& \\
T18& M. Person, A. Arredondo&                          $-$31.780278& $+$18.622902&   35&  6.0&  2816& \\
T19& J. Moore, S. Strabala&                            $-$31.286389& $+$23.699167& 1287&  6.4&  2898& \\
T20& T. Blank, P. Maley, H. Throop, N. Erasmus&                    $-$31.046868& $+$22.992324& 1272&  4.1&  1780& \\
T21& A. Verbiscer, A. Caspi, T. Ruhland&               $-$32.564777& $+$18.977851&  203&  6.0&  3222& clouds at the end\\
T23& M. Nelson, P. Hughes&                             $-$30.713013& $+$23.904314& 1241&  2.1& \nodata& \\
T24& B. Andersen, J. Wilson&                           $-$30.618650& $+$22.897806& 1187&  5.9   & 16     & \\
T25& M. Skrutskie, D. Josephs&                         $-$30.670305& $+$23.567372& 1193&  1.8& 2380     & \\
\enddata
\tablecomments{\scriptsize
Positions are all referenced to WGS84 datum.
}
\end{deluxetable}

\subsection{Observations}

Twelve stations in each continent successfully deployed and all collected useful
data.  Table~\ref{tbl-jun3sta} provides a summary of the mobile stations.   All
Earth-based positions for this deployment are provided on the WGS84 datum.  Every
station in Argentina had clear conditions but some had to deal with preventing
formation of dew on the telescope optics.  The teams near Clanwilliam, South
Africa had variable amounts of clouds, but the teams that headed east had clear
skies.  Because the Moon had set in South Africa, the teams in South Africa
experienced systematically lower background levels.  They also had better
seeing than the teams in Argentina.  The Argentina teams observed with a 66\%
illuminated Moon and higher contributions from light pollution, which resulted
in generally higher background noise levels.  The background information for T23
is not provided due to it being a very different system and the inter-comparison
with other stations is not particularly useful.  All stations used a 0.5-sec
exposure time.  The shadow speed of 20~km/sec meant a central chord on a D=40~km
body would be four frames.  We ran all observations for 45 minutes centered on
the local predicted event mid-time.  We designed this range of time to cover
the stable region of the estimated Hill sphere for \musn.  We did not see any
lightcurve signatures related to \musn\ in any of the data sets -- fixed or mobile.
Figures~\ref{fig-lc0603a} and \ref{fig-lc0603b} show the data from the mobile
stations.    These figures only show data within 30 seconds of the predicted event
mid-time.  The lightcurves are sorted north to south across the predicted track.
There is a lot of variability in data quality as can be seen in the plots.  Most
of the apparent dropouts in these data are due to high winds and severe image
smearing.  In these cases, we visually examined the data to confirm that the target
star was, in fact, still visible and the dropout was not an occultation.

\begin{center}
\includegraphics[scale=0.5]{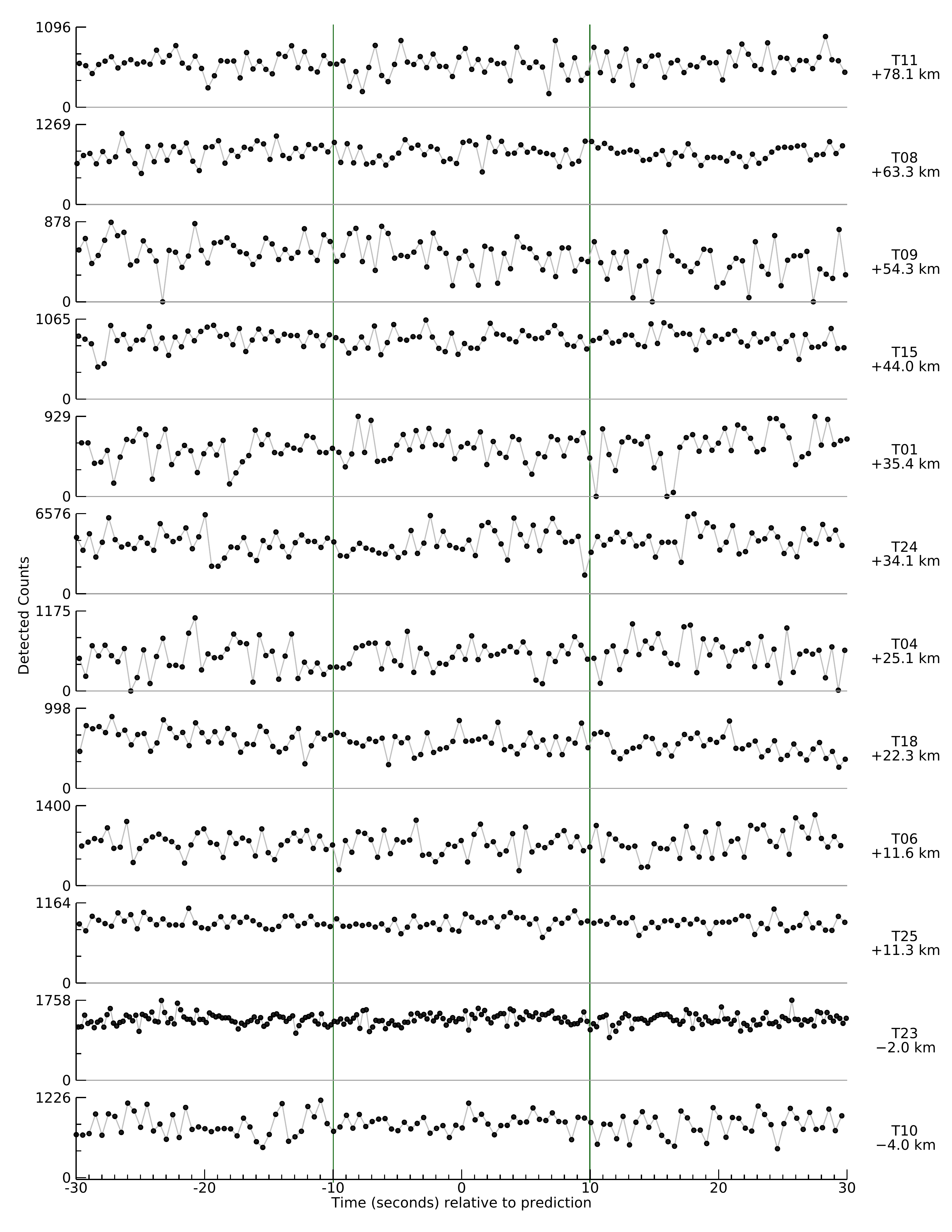}
\figcaption[figlc0603a.pdf]{\label{fig-lc0603a}
Observations from 2017-06-03 occultation, part 1.
The figure shows the lightcurves from the northern half of the data
collected by the mobile stations.  Each sub-plot is labeled
on the right with the team number and the cross-track offset.  The team numbers
are cross-referenced with Table~\ref{tbl-jun3sta}.
The plots indicate the signal level from each station -- higher numbers
indicate higher signal levels.  The green vertical lines indicate
the predicted 3-$\sigma$ uncertainty limits for the event.
An electronic copy of the data in this figure is provided.
}
\end{center}

\begin{center}
\includegraphics[scale=0.5]{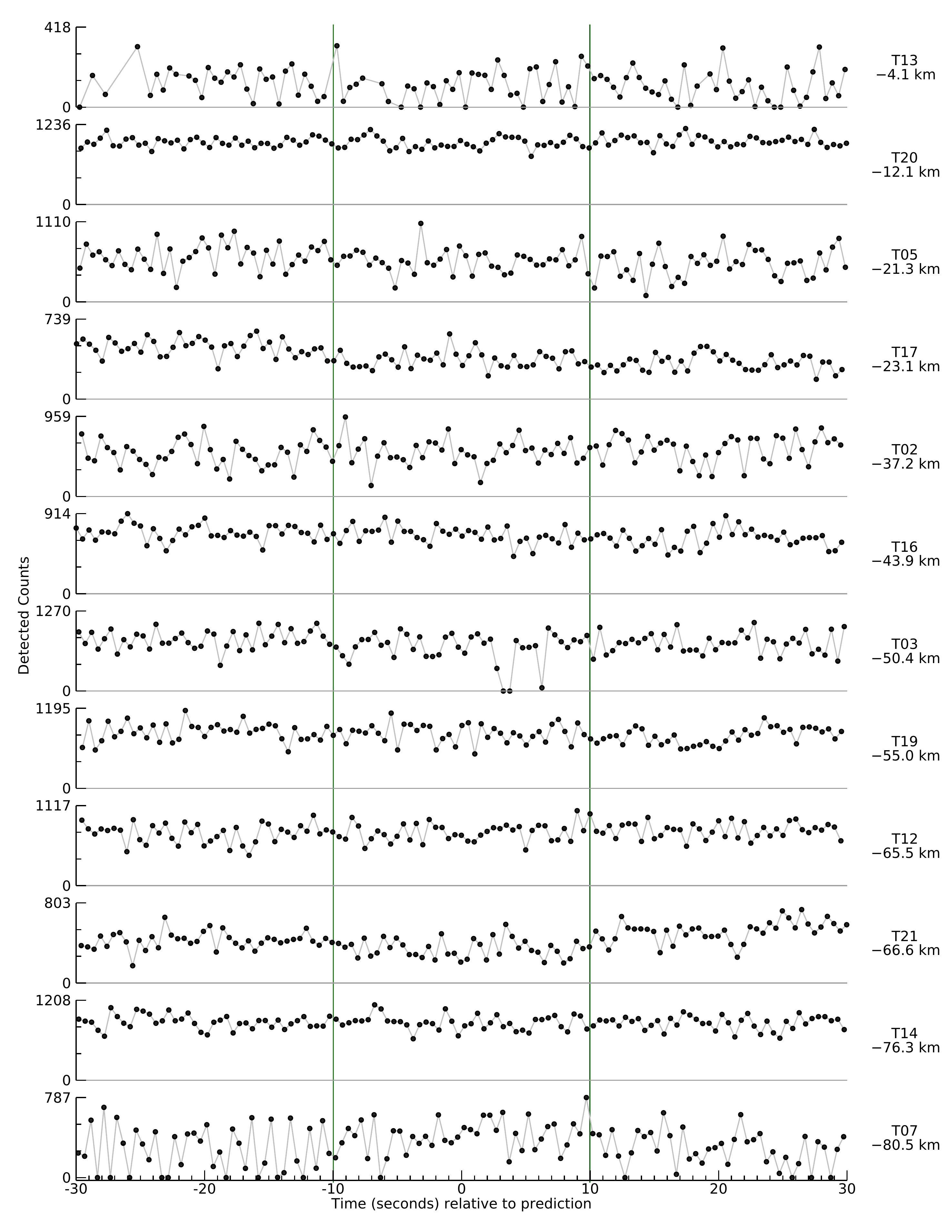}
\figcaption[figlc0603b.pdf]{\label{fig-lc0603b}
Observations from 2017-06-03 occultation, part 2.
The figure shows the lightcurves from the southern half of the data
collected by the mobile stations.  Each sub-plot is labeled
on the right with the team number and the cross-track offset.  The team numbers
are cross-referenced with Table~\ref{tbl-jun3sta}.
The plots indicate the signal level from each station -- higher numbers
indicate higher signal levels.  The green vertical lines indicate
the predicted 3-$\sigma$ uncertainty limits for the event.
An electronic copy of the data in this figure is provided.
}
\end{center}

\needspace{5\baselineskip}
\subsection{Fixed Stations}

\begin{deluxetable}{llP{4cm}ccccl}
%\footnotesize
\tablecaption{Fixed Observing Stations\label{tbl-fixed}}
\tablewidth{0pt}
\tablehead{
\colhead{Name}&
\colhead{Event}&
\colhead{Team}&
\colhead{Latitude}&
\colhead{E Longitude}&
\colhead{Elevation}&
\colhead{X-track}&
\colhead{Comments}\\
& & &
\colhead{(deg)}&
\colhead{(deg)}&
\colhead{(m)}&
\colhead{(km)}
}
\startdata
Gemini&      MU20170603& W.~Fraser&     $-$30.240750&  $-$70.736693& 2722& 384& \\
SAAO&        MU20170603& A.~Sickafoose, A.~Genade& $-$32.378944& $+$20.811667& 1760&  17& \\
EABA&        MU20170603& M.~Santucho, E.~Pulver, H.~A.~Durantini Luca, R.~Artola &$-$31.568442& $-$64.549836& 1350&\\
C\'ordoba&    MU20170603& C.~Colazo, R.~Melia&$-$31.599167 & $-$64.548333 & 1350 & & \\
SARA-CT & MU20170603 & A. Bosh & -30.17200833 & -70.79916667 & 2012 &  & observing through clouds \\
\hline
Gemini&      MU20170710& W.~Fraser&     $-$30.240750&  $-$70.736693& 2722& & \\
SOAR&        MU20170710& A.~Zangari, L.~Young, J.~Carmargo&$-$30.237892&$-$70.733611& 2748&  & \\
IRTF& MU20170710& S. Benecchi & $+$19.8262 & -155.4719 & 4205\\
\hline
SOAR&        MU20170717&L.~Young, J.~Carmargo&$-$30.237892&$-$70.733611& 2748&      & \\
El Leoncito& MU20170717&   E.~Garc\'ia-Migani,R.~Gil-Hutton        &     $-$31.798600&  $-$69.295600& 2483&      & very bad seeing \\
duPont & MU20170717 & A. Bosh & -29.01583333 & -70.69194444 & 2380 &   & good weather\\
\hline
\enddata
\tablecomments{\scriptsize
EABA = Estaci\'on Astrof\'isica Bosque Alegre, C\'ordoba
}
\end{deluxetable}

\subsubsection{Gemini}
We acquired observation data at the Gemini-South telescope on Cerro Pach\'{o}n
using the Gemini Acquisition camera (AC) and a similar methodology as
\citet{fra13}. The AC is a shutter-less 1k$\times$1k frame transfer CCD camera
with pixel scale of 0.12$''$/pixel that supports sub-frame windowing. We acquired a
nearly 60-minute sequence centered on the nominal overhead passage time. We
positioned the CCD so that the target star and a nearby reference star were
fully included in the window.  The CCD was read out with 2$\times$2 on-chip binning
with a window of 88$\times$65 binned pixels. We used an exposure time of 0.1~s. 

Nominally, the Gemini header creation system creates image timestamps. However,
this system was never intended to operate at the high cadences of our sequence.
Because of this limitation, we disabled the header creation system to maximize cadence
and minimize inter-exposure deadtime. We created image timestamps by monitoring
file creation times which were produced by GPS time within the Linux system.
This imaging configuration resulted in a 0.107~s median deadtime due to image
readout and file writing.  The resulting image cadence was 4.8~Hz.

We de-biased and flattened science frames in the usual manner using sky-flats.
We extracted photometry using the SExtractor software package, and calibrated
the relative flux of the target star using the brighter reference star. The
resulting photometry had a mean signal-to-noise ratio (SNR) of 48.  No occulting structures, dust, rings,
or solid bodies were seen.  The cross-track offset for these data was 384 km,
too far away to be relevant for the solid-body occultation.

\subsubsection{SAAO}

We also took observations on the 74-inch telescope at the South African Astronomical
Observatory (SAAO) using one of the Sutherland High-speed Optical Cameras
\citep[SHOC;][]{cop11}.  This instrument is optimized for stellar occultation
observations utilizing a frame-transfer CCD which can trigger each image from a
GPS.  The conditions on the night of the event were good, with scattered, light
clouds and seeing of roughly 1.4 arcsec.  For these observations, we took 27000
frames starting at 02:47:00.0 UT with a cadence of 0.1~s and a 6.7-msec deadtime.
We set the instrument to $-70$~\mydeg C in 3~MHz conventional mode with the
5.2$\times$ amplifier, binned 8$\times$8 (for a plate scale of 0.608~arcsec/pixel),
and no filter.

We reduced the data using biases taken on the night of the event and flat fields
taken the previous night, which was cloudless.  We performed photometry on the
target star and the one nearby brighter comparison.  We carefully selected a
background region to avoid other stars in the field. The optimal aperture was 6
binned pixels or 3.65 arcsec.  Figure~\ref{fig-lcsaao} shows the resulting
differential light curve, normalized to one, with a signal-to-noise ratio (mean
over standard deviation) of 21.  These data are the closest to the shadow
centerline and have significantly higher SNR and time resolution compared to the
mobile stations.  We saw no evidence for any solid body  event.  The data have a
cadence of roughly 2~km per sample and grazing events as short as 200~m can be
ruled out.  In the subsequent analysis, we simply treat this as a non-detection
and do not consider potential grazing chord constraints.  We will return to the
constraints provided by these data when discussing the data from all occultation
events together (see \S\ref{sec-combine}).

\begin{center}
\includegraphics[scale=0.5]{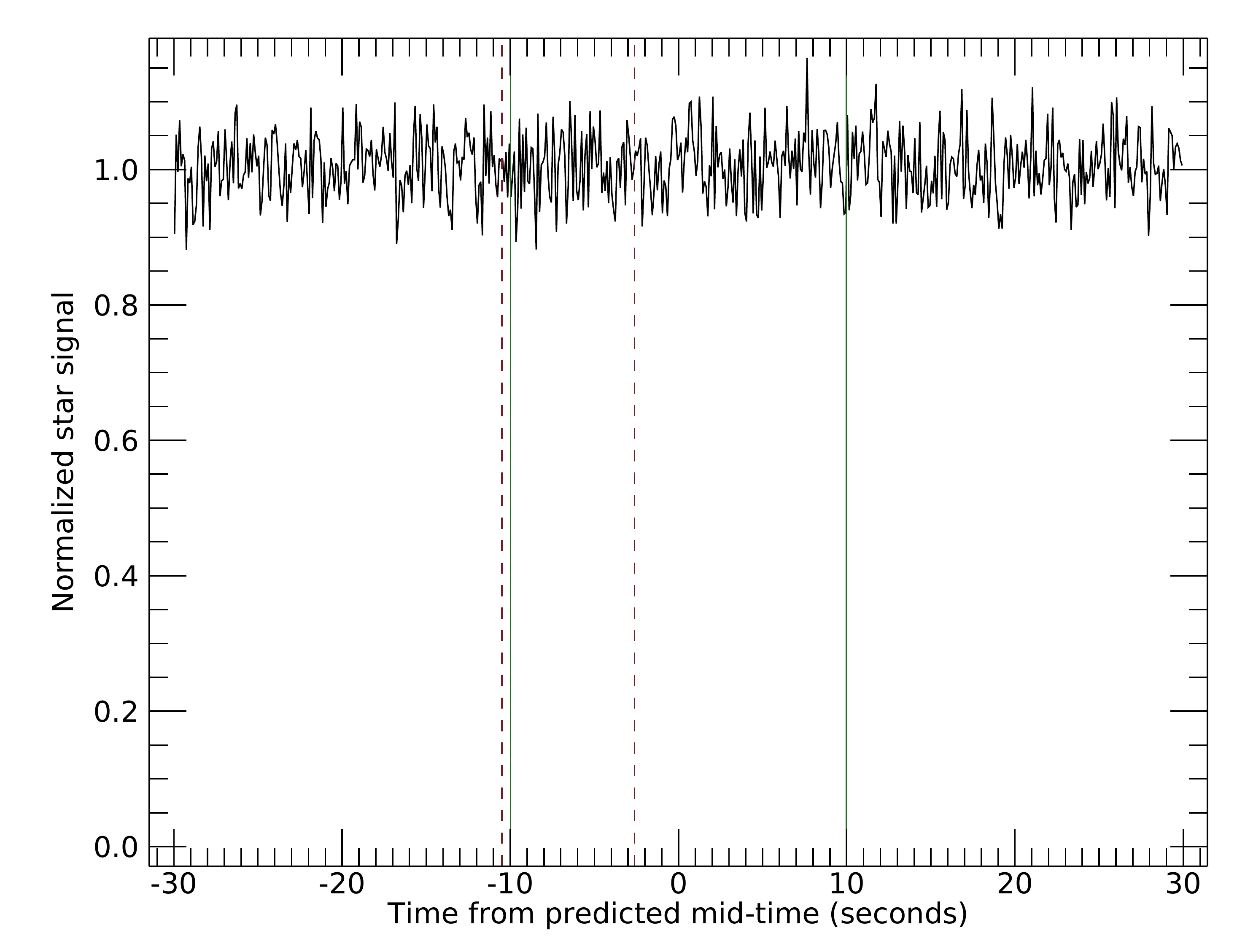}
\figcaption{\label{fig-lcsaao}
SAAO observations for the 2017-06-03 occultation.
The figure shows the lightcurve obtained from SAAO.  We see no event in the
data.  The solid green lines indicate $\pm 3 \sigma$ relative to the ``may25a''
prediction.  The dashed purple lines indicate $\pm 3 \sigma$ relative to the
final ``ey7'' post-diction.  The cross track offset based on ``ey7'' is
17.4$\pm$4.0~km.  See Section~\ref{sec-combine} for an explanation of the terms ``may25a'', and ``ey7''.  An electronic copy of the data in this figure is provided.
}
\end{center}

\subsection{Results}

Based on the final prediction prior to the June 3 event and the actual
site locations, the chance of a getting zero chords for D=40~km was 3\%.
For a D=40~km object, a single chord would be based on multiple frames with
the star occulted and could be recognized with high confidence.  For D=20~km,
the chance of a null result was only 14\%.  In this case, either zero or one
chord would likely be a null result since the chord would be so short. It was very
unlikely that we would have seen even smaller objects because the chance of
getting a single chord at all was so small.  Even if we got a chord, there was
a large chance it would be too short to be recognizable. We also did not observe a
solid-body event from any fixed site; however, most fixed sites were too far
away from the mobile chords to provide much constraint.

Under the assumption that all of our error sources (random and systematic) are
known and well characterized, these results indicated the large and dark case for
\musn\ was unlikely.  We chose to then optimize for the small object case for
subsequent occultations.  

\subsection{Limits on moons and opaque rings from Gemini\label{sec:dust}}

The non-detection of any occultation in the Gemini data provided upper limits on
the presence of dust particles within the Gemini beam. The inter-exposure deadtime
was the limiting factor in the size of detectable particles in the \musn\
environment.  The minimum detectable size was a particle perfectly centered on the
occulted star, which would have cast a shadow on the detector that would be mostly
contained in the deadtime, but with just enough on-exposure shadow to cause a
detectable dip in flux. We consider a 5-$\sigma$ dip such that we would not have
expected any one of the $\sim$16,000 exposures to vary by this amount by chance.
Thus, in this limiting case, a shadow of duration $t = \frac{5}{SNR}\,t_e + t_d $
where $t_e=0.1$~s was the exposure time, and $t_d=0.107$~s was the deadtime, could
have produced a detectable dip in flux. With a ground-track shadow velocity of
20~km~s$^{-1}$, the minimum detectable particle size (or narrow and opaque
ring) was 2.3~km.

\section{2017 July 10 Event}

We originally considered this event for a large mobile deployment.  Because of a
number of logistical difficulties, we focused on large aperture observations to
search for or constrain the presence of rings or diffuse dust structures.  The
global view of the ground track in Figure~\ref{fig-0710globe} shows some of the
difficulties.  The shadow only crossed land in regions of South America that were
unlikely to be clear.  More importantly, this event occurred just 17\mydeg\ from a
99\% illuminated moon and the occulted star was the faintest star of the four.
The largest effort went into supporting an observation with SOFIA \citep{tem14}.
Data were collected at other fixed sites but in the end were less constraining
than the Gemini data from 2017-06-03.

\begin{center}
\includegraphics[scale=1.0]{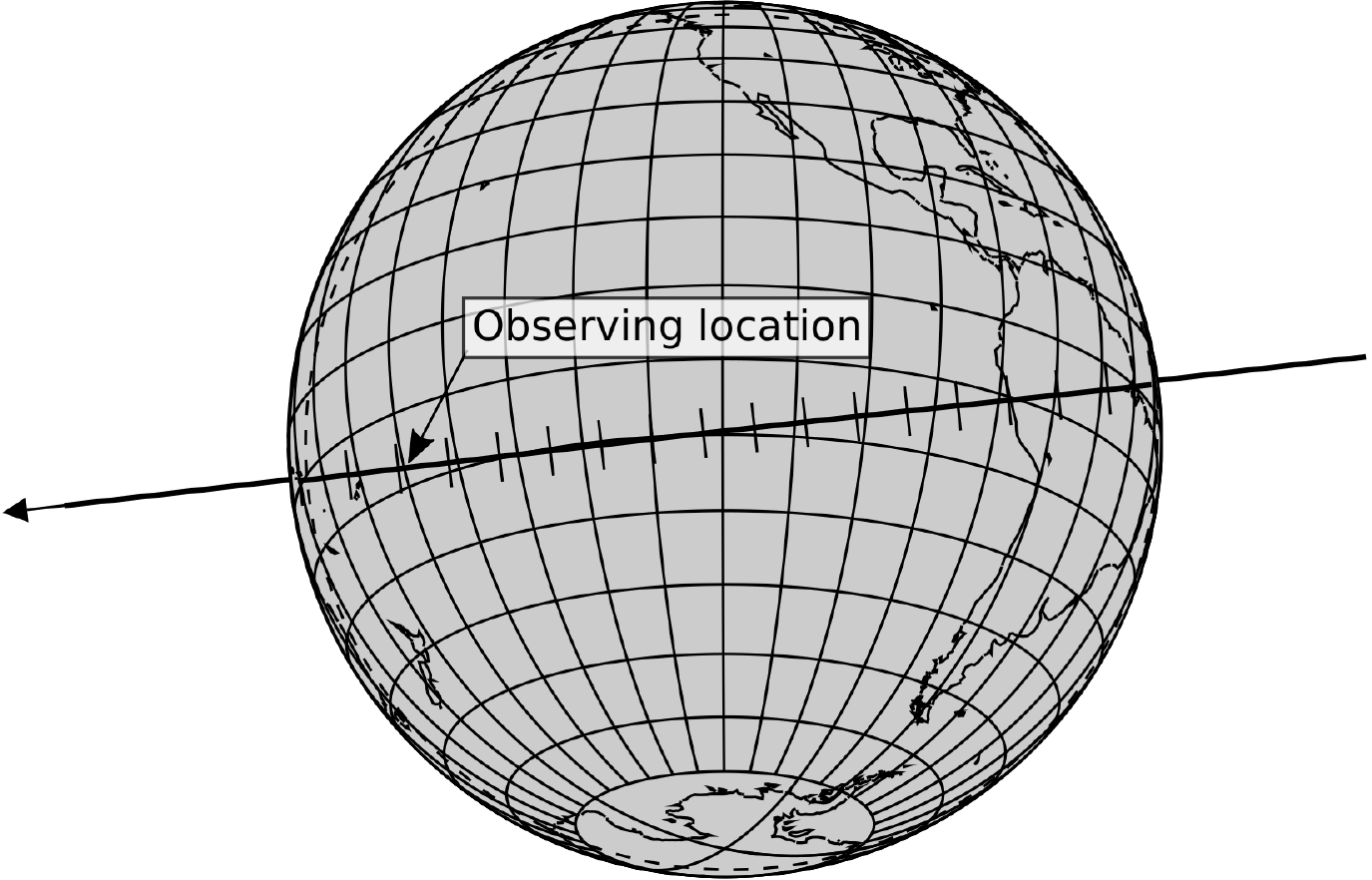}
\figcaption{\label{fig-0710globe}
Global view of 2017-07-10 occultation prediction.
The figure shows the Earth as seen from \musn\ at the time of
geocentric closest approach.  The Sun is below the horizon in the
regions shaded gray (the entire globe).
The dashed line shows where the Sun is at -12$^\circ$ altitude.  The solid line indicates the predicted ground-track with
the width drawn to scale for a 30 km diameter object.  The arrowhead indicates the direction of motion and the ticks are spaced at
30 second intervals from 07:41:30 to 07:50:00 UT.  Shadow velocity was
25.0 km s$^{-1}$.  A 99\% illuminated Moon was 17\mydeg\ away from the target
at the time of the event.
}
\end{center}

\subsection{Prediction}

SOFIA approved this flight opportunity because the likelihood of a positive
outcome was deemed sufficient.  Of the three 2017 events, it was the only one
with a track accessible using a single flight from the summer deployment base
of Christchurch, NZ.  The primary goal of the flight was to probe the system
for material potentially hazardous to New Horizons.
A secondary goal was to observe a solid body event; however, it was very
challenging to target the aircraft on the occultation track.
The support requirements for the SOFIA flight set the timing of the prediction
work for this event. 

A large observing campaign with HST on \musn, timed to be completed just prior
to the SOFIA flight, significantly improved the prediction.
HST GO-14627 (PI: Benecchi) provided
24 orbits that returned 119 images and resulted
in 118 new astrometric measurements during the interval from 2017 Jun 25 to
2017 Jul 4.  The goals of the lightcurve investigation dictated the timespan
and spacing of the observations, while we set the end time of those
observations to allow all of these data to be included in the SOFIA
prediction, with the smallest temporal gap between the end of data and the
time of the occultation opportunity.  The new dataset doubled the number
of astrometric measurements between this and the previous occultation.
We once again reduced all images against the pre-release version of the Gaia DR2
catalog \citep{gai18}.  The vastly improved prediction uncertainty for the
cross-track position was 14~km.  The in-track error (timing) was 21~km
(0.86 sec).

\subsection{Deployment}

Assuming no error in delivering SOFIA to the target point, the odds of
getting one chord on a D=40~km object were 76\% based on the final prediction.
As the object size
decreases the odds of getting a chord decrease.  Even at D=20~km,
the chance of getting a chord was 44\%.  The targeting window for the 
flight was to place the aircraft
within 1~km and 1~sec of the aim point along the track (due north of
New Zealand).
We chose a target point on the shadow centerline of 2017~Jul~10 07:49:11 UTC at latitude $-$16.403333\mydeg\
and east longitude of 184.960000\mydeg.

%JD 2457944.82582176
% from geom_lc1.dat
% 2457944.82582161 -16.398726 184.957986 12354.7
% 2457944.82582219 -16.398831 184.957959 12354.7

\subsection{Observations}

\subsubsection{SOFIA}

SOFIA contains a Focal Plane Imager (FPI+) based the Andor iXon
DU-888 commercial CCD camera \citep{pfu18}.  The FPI+ usually serves as the guide camera for SOFIA,
receiving visible wavelengths via the telescope’s dichroic tertiary mirror.
Many of the FPI+ characteristics that are advantageous for a guide camera (fast
read-out rates, zero dead time between frames, high quantum efficiency and low
read noise) are also critical for an occultation camera, where the desired cadence
often produces scenarios with low source counts \citep{pfu16}.  In this case, given the relative
velocity between \musn\ and the occultation star of 24 km s$^{-1}$, we planned for a
sampling rate of 20~Hz. That rate was intended to detect rings with equivalent
widths of $\sim$1~km.

Our SNR estimator predicted that each 0.05-s exposure would have an SNR of 13,
assuming an open filter, an occultation star G-mag of 15.57 and 4$\times$4 pixel binning
(for an effective plate scale of 2.04$''$ per spatial element). Our SNR estimate was
comparable to the published FPI+ sensitivities (SOFIA Observers Handbook, Figure 
5.1), where an SNR of $\sim$50 is expected in a 1-s exposure of a 15.6 V-mag star.
Unfortunately, the full Moon was fewer than 10 degrees away from \musn\ during the
event, and the highly variable background counts became the dominant noise source. 
In practice, we found that the SNR per timestep was between 3.5 and 5. Nevertheless,
the SOFIA/FPI+ lightcurve produced the first occultation detection of
\musn\ --
a very short grazing chord, though this was not clear until much later.

Based on the final aircraft telemetry, we were 550~m
from the target point at the target time.  The minimum distance from the
target point was 1.8 seconds later at a distance of 330~m.  This degree of
success in getting to our aim point took considerable skill and effort on the
part of the flight crew and demonstrates what SOFIA can achieve
despite the lack of tools for this specific purpose.  The most difficult
requirement levied on the flight was getting to the
aim point at the right time.  We attempted to get within 1 second and got very
close.  In the end, the limiting aspect of our deployment was our ability
to predict the right place.

We saw no obvious signs of an occultation during the flight on the real-time
monitors.  The data required very careful photometric extraction because
individual images had rather low apparent SNR on the target star.
Figure~\ref{fig-lc0710} shows one of the five independent lightcurve reductions.
In these data, there is one singularly deep dip in the lightcurve at about 45~km
prior to minimum separation.  This particular analysis is the result of
2$\times$ image binning prior to photometry.  We did not apply any spatial shifts to the
images and combined the images prior to processing.  Because of how the images
are indexed, there were two possible binned outcomes.  Here we show the outcome
that returns the strongest dip. The other option shows a weaker two-point dip.
Looking at the original frames prior to combining, the star is missing on the
frame at the center of the dip while its flux is reduced somewhat on the frame
before and the frame after the center.  At our sampling
rate, this dip corresponds to a solid-body chord length of $\sim$1~km.  We did
not immediately identify this dip, but independent processing of the data
confirmed the short dropout.  Because on-chip reference stars do not show
this dip, we
believe this to be a real signature associated with \musn.  However, we did not
complete this analysis until after the third deployment.  As far as we knew at the time
of the July 17 event, we had come away empty-handed from the first two attempts.

\begin{center}
\includegraphics[scale=0.5]{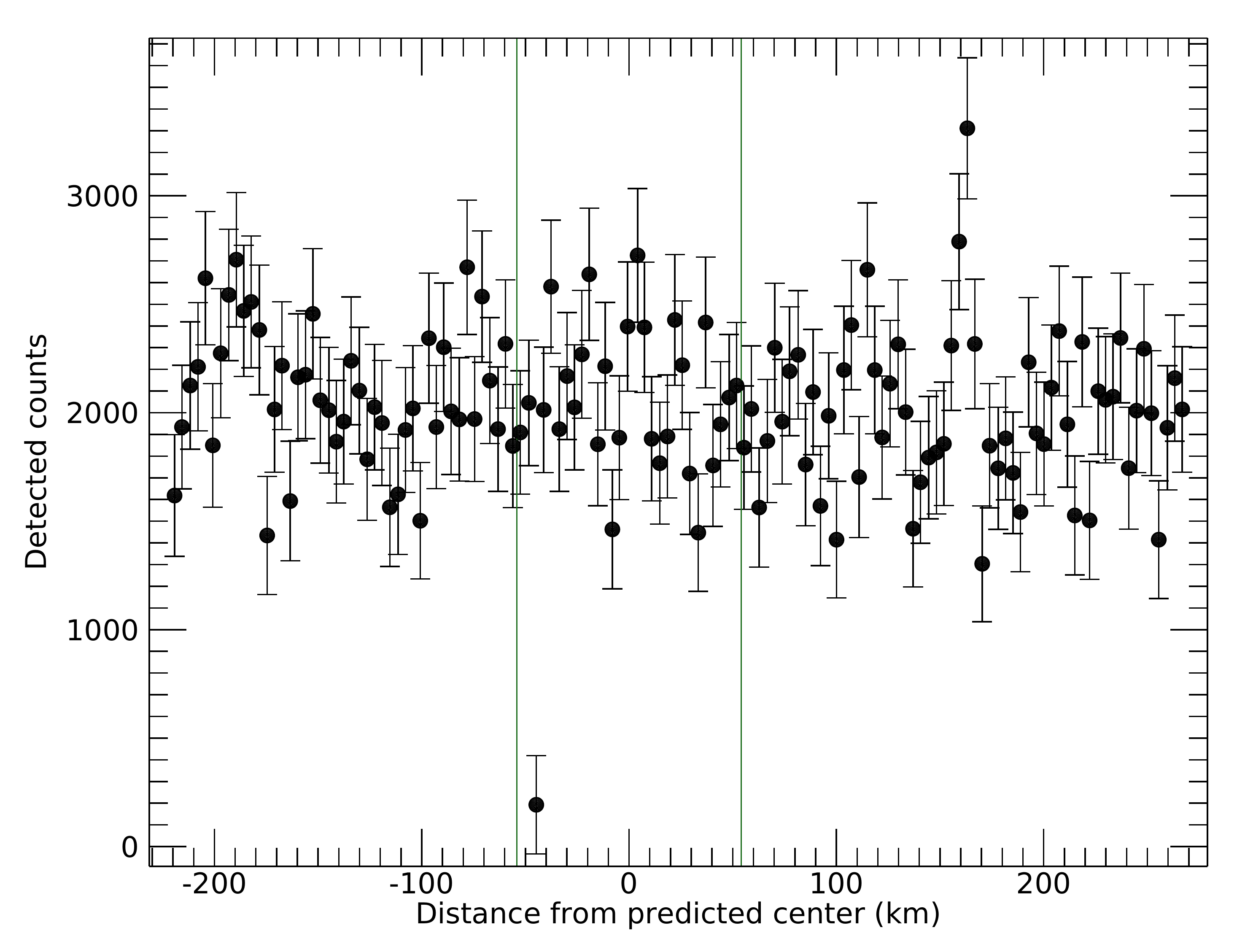}
\figcaption[figlc0710.pdf]{\label{fig-lc0710}
Observations from the 2017-07-10 occultation with SOFIA.
The figure shows the lightcurve with a single-point dropout about 45 km
prior to the predicted minimum separation.  The green vertical lines indicate
the predicted 3-$\sigma$ uncertainty limits for the event.
An electronic copy of the data in this figure is provided.
}
\end{center}

With respect to timekeeping during the flight, we emphasized with both the
observing team and flight crew the importance of unambiguous timing information
for all images during the occultation sequence. We carried out numerous tests
during the outbound flight leg prior to the occultation.  We identified and
compared four different time sources during flight: 1) navigation clock used on
the flight deck, 2) clock at the science flight control station, 3) GPS-slaved
time source used by FPI+ recorded with the science images, and 4) GPS time from
an application running on a cell phone with GPS receiver.  There was no way to
electronically measure the differences between these time sources and we had to
rely on verbal callouts of the clocks to investigate offsets.  Past experience
with this type of test shows that one can, with care and practice, detect shifts
down to about 0.1~seconds.  Clock \#4 proved to be unreliable with variable
shifts compared to the other three and we discounted this time source.  None
of the other three clocks were identical.  Clock \#1 was 1 second ahead of
Clock \#2 and Clock \#2 was 1 second ahead of Clock \#3.  After careful analysis
of the various systems, we concluded that Clock \#3 was the one that had been
most heavily tested and best understood and were confident that the time tags
on the images were correct to within the usual precision limits
of a GPS-based system.  The targeting of the aim point in time may have been
affected by these clock offsets though there was no meaningful degradation of the
experiment as a result.  A reasonable explanation for the offsets is that Clocks
\#1 and \#2 had (different) out-of-date leap-second almanac information that had
never before been recognized since they are not normally tested or relied upon to
this level.  We had a hard time testing this potential timing concern with the
occultation prediction uncertainties at the time.  The final post-event
reconstructions show that our chosen time reference decision is clearly the most
reasonable giving us additional confidence that the FPI+ timing is correct as
expected.

\subsubsection{Gemini-South}

We acquired data with the AC on Gemini and reduced in a nearly identical fashion
to the June 3rd event reduction. However, we used a larger 120$\times$120 binned pixel
window to include a reference star in the frame, resulting in a longer mean
deadtime of 0.178~s and an effective imaging cadence of 3.6~Hz. We acquired a
45-minute imaging sequence centered on the nominal shadow passage time. The
effective photometric SNR of the science target was 17.  We did not detect
any occultation event in the Gemini data.  We computed minimum
detectable size limits as done in Section~\ref{sec:dust}.  Due to the increased
deadtime and lower SNR for this sequence, we found a larger minimum detectable
size of 5.39~km.

\subsubsection{SOAR}\label{sec-soarjul10}

We took data remotely at the SOAR telescope on Cerro Pach\'{o}n using
a Raptor photonics Merlin EM247 frame transfer CCD camera.  The Raptor
Merlin is a 658$\times$496 pixel CCD camera with 10 $\mu$m pixels spanning a
relatively narrow $\sim$60 arcsec by 60 arcsec FOV on the SOAR 4.1-m telescope. The Raptor
includes GPS-based timing.  We took images continuously through an open
filter wheel at a cadence
of 500 ms per image from UT 2017-07-10 06:53:39 to 2017-07-10 08:45:39,  and the signal-to-noise ratios near the middle of the observation window was 26.  This station was 2310 km from the shadow centerline.  No
solid-body, rings, or diffuse occulting structures were seen in the data.

%Observers: Amanda Zangari, Leslie Young, Julio Carmargo, Bruno Quint

% we are not sure of the FOV.  MY calculations using SOAR's number of 3.025 "/mm (http://www.ctio.noao.edu/soar/content/soar-technical-specs) and Raptor's sheet of 6.58 mm chip (http://extranet.on.br/camargo/GUGA/ManualRaptor/MerlinConfig/SP-ME247_V04_RaptorMerlin.pdf) gets us 19.9" and 15.0" for the FOV, which is two small based on looking at actual Merlin data,

\subsubsection{IRTF}

Although the 3.0-m NASA Infrared Telescope Facility (IRTF; Mauna Kea, HI) was not near
the centerline of the occultation prediction, we observed from here to look for
extended rings or moons in the vicinity of \musn.  S.~Benecchi with
assistance from S.~J.~Bus and A.~Zangari took data remotely. We used the MIT Optical Rapid Imaging System (MORIS), an
Andor iXonEM+ DU-897 camera co-mounted with SpeX (which we did not use) in
conventional readout mode.  We operated without a filter to get the
scale of 0.11 arcseconds and a field of view of 1 arcmin square
\citep{gul11}. Our exposure time was 0.5 seconds and we collected
3000 images over 49.4 minutes from 2017-07-10 07:16:50 until 18 minutes past the
predicted occultation time (UT 7:47). We stopped collecting data when clouds
claimed the sky, obstructing the field. The seeing was 0.8 arcseconds and SNR
on the occultation star in a single exposure was about 4.  Analysis 
of the lightcurve does not show any extended structures around \musn. 

\section{2017 July 17 Event}

This star was the brightest of the three candidates for 2017.  We could observe it
from the ground in southern South America.  A deployment for SOFIA was ruled out
because it required an extremely long double-length flight path.  In the end,
we concentrated all of our resources into a mobile deployment from a single location.
We ruled out a deployment to Chile because of the weather prospects at that latitude.
We determined that Southern Argentina was the most promising location even though
we were not certain we would have clear conditions. Climatic indications for the
area indicated a $\sim$50\% chance of workable conditions, good enough for the attempt.

\begin{center}
\includegraphics[scale=1.0]{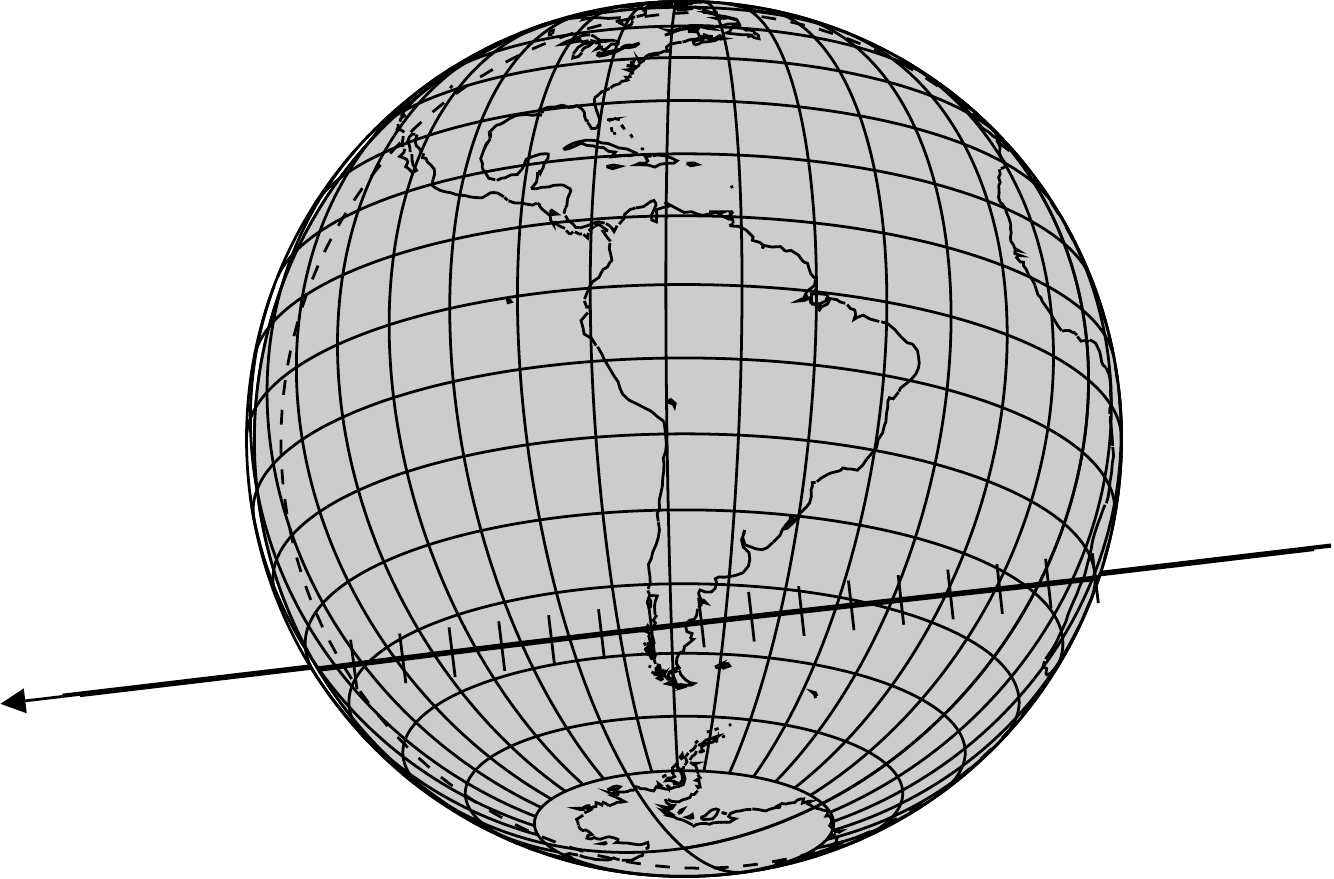}
\figcaption{\label{fig-0717globe}
Global view of 2017-07-17 occultation prediction.
The figure shows the Earth as seen from \musn\ at the time of
geocentric closest approach.  The Sun is below the horizon in the
regions shaded gray (the entire globe).
The dashed line shows where the Sun is at -12$^\circ$ altitude.  The solid line indicates the predicted ground-track with
the width drawn to scale for a 30 km diameter object.  The arrowhead indicates the direction of motion and the ticks are spaced at
30 second intervals from 03:46:00 to 03:53:30 UT.  Shadow velocity was
23.5 km s$^{-1}$.  A 46\% illuminated Moon was 105\mydeg\ from the target
at the time of the event.
}
\end{center}

\subsection{Prediction}

We based the final ground track prediction on the same ephemeris used for the
July 10 event.  Our uncertainty estimates gave a 1-$\sigma$ cross-track error
of 14.2 km (0.46 mas) and a timing error of 0.88 sec (0.70 mas).  These
uncertainties were higher than for July 10 due to a larger uncertainty in the
star position from Gaia DR2 \citep{gai18}.  Even so, this prediction was
three times better than that for June 3, permitting us to deploy ground stations
with smaller spacing.  As a precaution against a miss as on June 3, instead of
deploying to an extremely tight spacing between stations, the
deployment plan covered just over 4-$\sigma$ in cross-track spacing relative to the
prediction, with a mean spacing between sites of 4 km.  Unlike the June event,
this coverage and spacing was sufficient for all plausible albedos.

\subsection{Deployment}

We chose Comodoro Rivadavia to be the central base for the deployment
based on local resources, commercial air carrier access,
proximity to a major road for transporting the equipment, and proximity to the
predicted ground track.  At the time this choice was made, the ground track was
still uncertain by 200-300 km.  The equipment that was used in South Africa was
sent via air freight directly to Buenos Aires where it was recombined with the equipment
used in Mendoza in June.  Everything was then trucked down to Comodoro Rivadavia
and distributed to the teams and their vehicles.

We built our deployment pattern around a 4-day schedule of events similar to the
June 3 deployment.  The first night all the teams gathered at the same location in
town to test the equipment. The first night provided all teams with an opportunity
to practice with the system with other teams nearby. This was particularly useful to
troubleshoot difficulties, especially for those new to the telescopes.  We split the
large group into four subgroups for the second night and sent them to different
nearby locations to practice a deployment with less help available.  We treated
the third night as a dress rehearsal, choosing site locations as if they were actual
locations.  This strategy provided us with an opportunity to test both the site
choices and the teams as if it were the actual event night.  
The fourth night was the event night.  Incredibly, all four nights were workable which allowed us every
opportunity for practice and event-night observations.

This deployment was not without challenges.  We had very few roads to work
with and our mobility was highly constrained.  This forced most teams to work
rather close to National Route 3 (RN3), the major north-south highway along the coast.
Stray light from passing vehicle headlights caused significant time-variable
glare on the images during data collection.  Also, this region of Argentina is
well known for its generally windy conditions.  Indeed, one of the nicknames for
Comodoro Rivadavia is the ``Capital of the Wind.''  The conditions started out very
calm on the first night but the wind gradually grew in strength with each passing night. The consensus
among the teams after the third night was that stray light and telescope shaking
from the wind would create serious problems for the event night.  The National
University of Patagonia, Comodoro Rivadavia campus, (Prof.~Marquez) and the
mayor's office offered an amazing amount of help with our difficulties.  We
received help from Prof.~Marquez in designing and building
windbreaks to shield the telescopes from the winds.  The mayor also offered to
provide large trucks as windbreaks.  We tested all of these on the dress rehearsal
night and found these made significant improvements in the data quality. The mayor
and the University suggested shutting down RN3 during the time of our
observations to address the stray light problem.  Local authorities enforced
a two-hour cessation of all traffic movement through the area where we
had deployed telescopes.  This level of help was absolutely essential to the
success of our efforts on event night.

Table~\ref{tbl-jul17sta} provides the final deployment locations.  We recorded all
of the positions from cell phone-based GPS applications and later confirmed
these measurements with Google Earth.  We experienced some variation in sky background
signal among teams but variation in image motion due to wind dwarfed the variation
in sky background signal.  Despite the wind mitigation efforts, the wind affected
all sites, though it affected some much more strongly than others.  Those
sites with high levels of wind shake had strongly varying image quality.  In the
end, this wind shake made the data reductions more difficult but not impossible.
Note that the seeing
values tabulated are really just an indication of the image quality for normal
quality images and the occasional large smearing from wind is not particularly
evident from the mean seeing tabulated.

\begin{deluxetable}{cP{5cm}cccccl}
%\footnotesize
\tablecaption{Mobile Observing Stations and Teams for 2017 Jul 17\label{tbl-jul17sta}}
\tablewidth{0pt}
\tablehead{
\colhead{ID}&
\colhead{Team}&
\colhead{Latitude}&
\colhead{E Longitude}&
\colhead{Elevation}&
\colhead{FWHM}&
\colhead{Sky}&
\colhead{Comments}\\
& &
\colhead{(deg)}&
\colhead{(deg)}&
\colhead{(m)}&
\colhead{(pixels)}&
\colhead{(counts)}
}
%\colnumbers
\decimals
\startdata
T01& A. Olsen, M. Dean&                      $-$45.918468& $-$67.606079&  181&  9.9&  1930& \\
T02& W. Hanna, P. Hughes &      $-$45.780608& $-$67.701558&  337&  9.3&  1909& \\
T03& C. Erickson, C. Wiesenborn, M. Camino&                              $-$45.838650& $-$67.838320&  384&  7.7&  1705& \\
T04& K. Getrost, L. Ferrario&                   $-$46.251567 & $-$67.608167&  27&  5.1& 1739& \\
T05& P. Tamblyn, R. Reaves&                            $-$45.710800& $-$67.360950&  47&  7.7&  1735& \\
T06& A. Rolfsmeier, T. Finley, C. Navarrete&                            $-$45.447329& $-$67.666419&  409&  7.5&  1759& \\
T07& S. Porter, B. Dean&                              $-$45.964509& $-$67.572241& 23& 6.9&  1894& \\
T08& D. Dunham, C. Ferrell, S. Makarchuk &                            $-$45.995417& $-$67.597194&  12&  7.7&  1689& \\
T09& T. Blank, K. Singer, Y. Kamerbeek&                            $-$45.889267&$-$67.779917&  220&  9.3&  1981& \\
T10& A. Friedli, D. Josephs &                  $-$45.216667& $-$67.233333&  587&  5.1&  1676& \\
T11& S. Conard, A. Resnick, P. Vidal&                 $-$46.318980& $-$67.582630&  5&  6.5&  1648& \\
T12& B. Keeney, A. Chapman&                            $-$46.206200& $-$67.624167&  8&  6.7&  1924& \\
T13& R. Venable, C. Lisse&                           $-$45.651700& $-$67.645600&     481&  9.8&  1880& \\
T14& S. Gurovich. S.A. Stern&                            $-$45.565300& $-$67.627500&     635&  7-12& \nodata&  drifted off target\\
T15& J. Moore, A. Lovell&                               $-$46.172500& $-$67.626944&    11 &  6.8&  1917& \\
T16& A. Verbiscer, C. Tsang, A. Daynes, F. Avelleros, G. Rotondo, I. Rotondo&                          $-$45.680692& $-$67.589490&   269  &  9.5&  1716& \\
T17& A. Zangari, C. Carter, P. Hinton, J. Fazio, M. Herrera&                           $-$45.529444& $-$67.618611 &    610 &  6.0&  1620& \\
T18& M. Buie, A. Ocampo, V. Saranitik&                          $-$45.823877& $-$67.460720&    11 &  8.2&  1964& \\
T19& J. Dunham, J. Mackie, P. Saizar&                            $-$45.484389& $-$67.633472&    472 &  4.4&  1734& \\
T20& A. Soto, J. Spagnotto, M. Pereyra&                               $-$46.104722& $-$67.628333&    7&  7.2&  1769& \\
T21& J. Jewell, S. Strabala, A. Heredia&               $-$46.062500& $-$67.624167&   7  &  6.9&  1894& \\
T23& M. Nelson, J. Skipper&                          $-$45.363655& $-$67.379478&   599 &    4--7& \nodata& variable seeing\\
T24& L. Wasserman, E. Golub, B. Dickason&                          $-$45.266111& $-$67.302500&  583&     \nodata&      \nodata& telescope failure\\
T25& M. Skrutskie, S. Henn&                         $-$45.307811& $-$67.329669& 589& 2.9   &  1760    & \\
\enddata
\tablecomments{\scriptsize
Positions are all referenced to WGS84 datum.
}
\end{deluxetable}

\subsection{Observations}

Twenty-two of the twenty-four stations were successful in collecting useful data
on the target star around the time of minimum separation. Table~\ref{tbl-jul17sta}
provides a summary of the data quality and notes for the mobile stations. The Moon
was 46\% illuminated and 105\mydeg\ away.  All stations used a 0.2-sec exposure
time.  We ran all observations for 45 minutes centered on the local predicted event
mid-time.  As with the other 2017 events, we designed this range of time to cover
the stable region of the estimated Hill sphere for \musn.  The T14 entry shows no
sky value and the data were also not processed due to the target star drifting off
the detector prior to the occultation and due to non-standard data collection
settings.  The T23 entry again shows no sky value due to it being a very different
setup and comparison of sky values has no meaning.  The T24 system suffered fatal
damage to its internal wiring and could not be repaired in time for the event.

\subsection{Data Reductions}\label{sec:2017datared}

We processed all the standard systems (T01-T22) data together, similar to the
June 3 data.  We copied images within $\pm$1~minute of the predicted mid-time out
of the full dataset for processing.  There was no measurable need for bias, dark,
or flat field calibration steps and therefore we did not apply these types of
corrections.  However, the raw images contained a low-level horizontal striping
pattern.  This striping is a feature of the bias pattern inherent in the detector
readout and varied from frame to frame.  We easily removed the pattern by computing
a robust mean for each row and then subtracting that mean.  Each image had a large
number of stars ($\sim$100) from which we generated a frame-by-frame
numerical point-spread function (PSF).  We then fit the PSF to all the stars by
adjusting the position and flux.  However, we excluded the target star from 
fitting at this step.  From the fit positions, we derived an astrometric solution
for each image based on the Gaia DR2 star catalog positions \citep{gai16},
corrected for proper motion to the epoch of the images.  We then used the
astrometric solution to compute a pixel location for the target star.  At this
point, we computed a PSF fit to the target star where the only free parameter was
the star flux.  We then computed the mid-time for each observation from the
GPS time and exposure time recorded in the header of each image.

This data processing methodology was quite valuable, particularly for the images
with significant image smear due to wind.  In these cases, the PSF is
arbitrarily complicated -- not just a simple linear smear.  As long as the PSF fit
accurately replicated the smear, it was possible for us to extract a useful target
star flux.  Taking full advantage of the PSF method required significant manual
effort to guide the PSF building and fitting process.  We only applied this extra
effort as needed to critical images in and around the time of the actual event.
Without this extra corrective step the target star may appear to drop out for a
frame or two.  We inspected these cases visually and saw a tortuous wind-driven
PSF; however, the target star is still visible.  For non-critical images, we noted
that the star is still visible and a given dropout is not interesting.

\begin{center}
\includegraphics[scale=0.55]{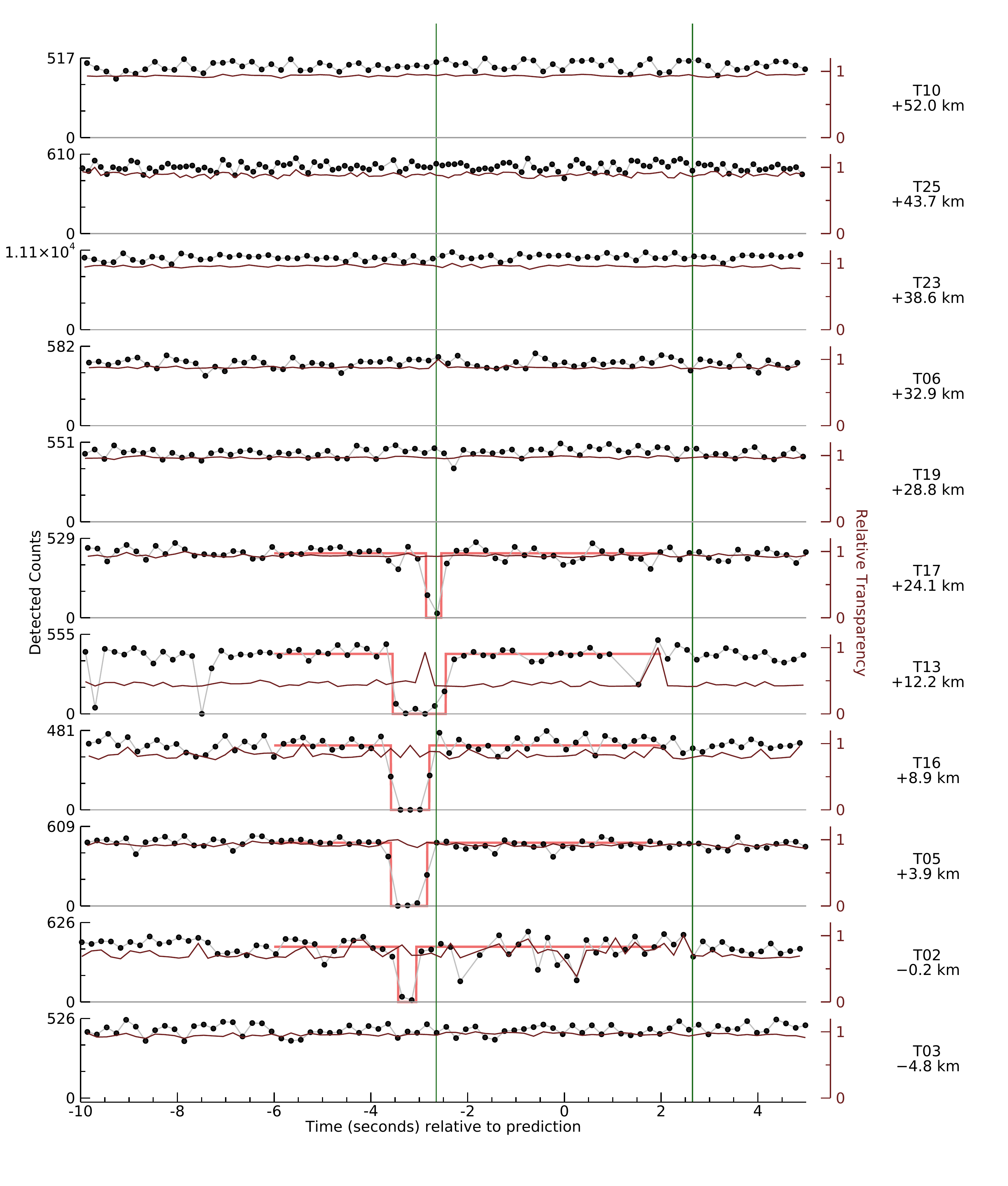}
\figcaption[figlc0717a.pdf]{\label{fig-lc0717a}
Observations from 2017-07-17 occultation, part 1.
The figure shows the lightcurves from the  northern half of the data
collected by the mobile stations.  Each sub-plot is labeled
on the right with the team number and the cross-track offset.  The team numbers
are cross-referenced with Table~\ref{tbl-jul17sta}.
The plots indicate the signal level from each station -- higher numbers
indicate higher signal levels.  The green vertical lines indicate
the predicted 3-$\sigma$ uncertainty limits for the event.
The second (brown) curve plotted is an estimate of the relative transparency.
Five of these curves show an overlain solid-body model used to extract
occultation timings.
An electronic copy of the data in this figure is provided.
}
\end{center}

\begin{center}
\includegraphics[scale=0.55]{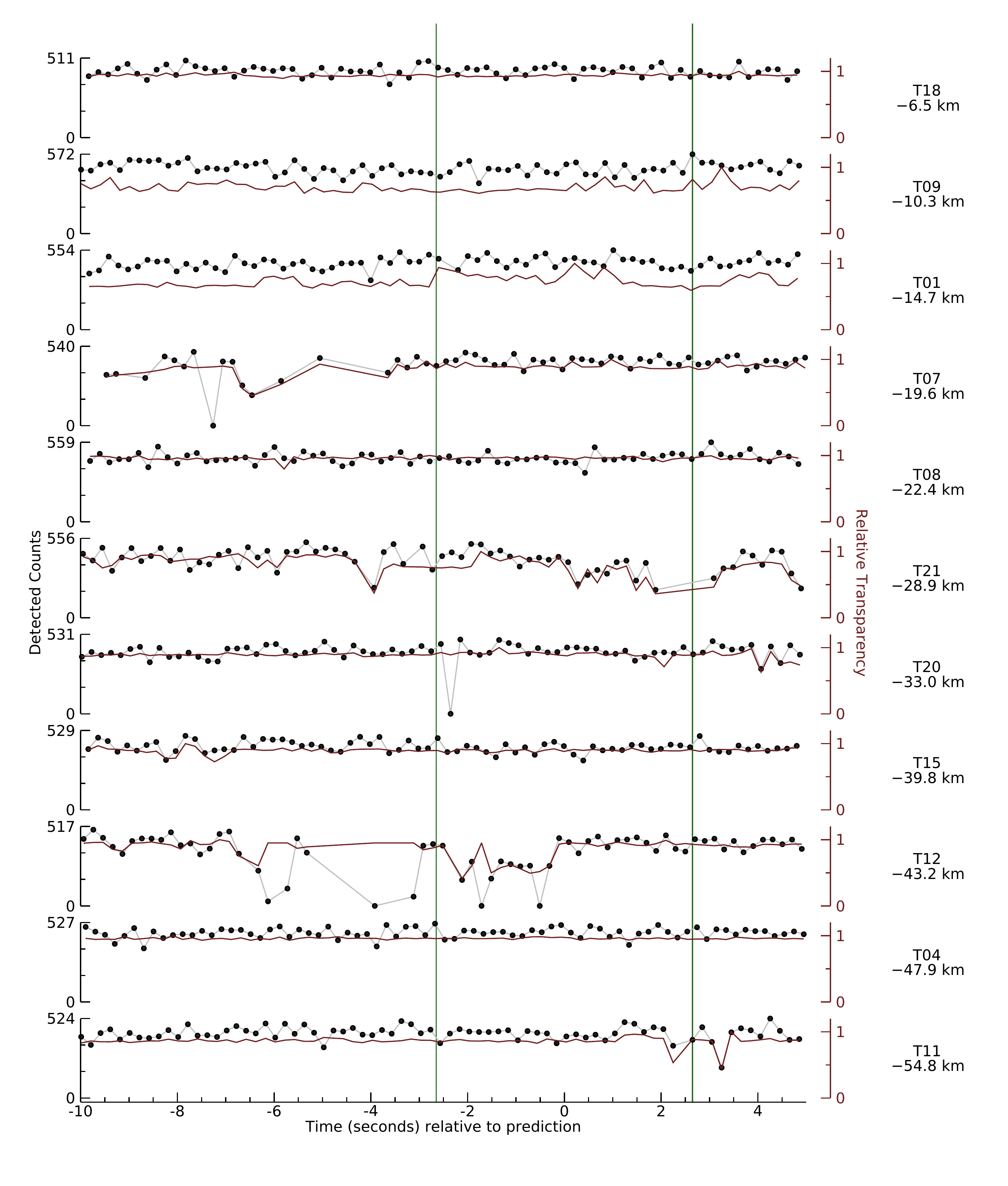}
\figcaption[figlc0717b.pdf]{\label{fig-lc0717b}
Observations from 2017-07-17 occultation, part 2.
The figure shows the lightcurves from the southern half of the data
collected by the mobile stations.  Each sub-plot is labeled
on the right with the team number and the cross-track offset.  The team numbers
are cross-referenced with Table~\ref{tbl-jul17sta}.
The plots indicate the signal level from each station -- higher numbers
indicate higher signal levels.  The green vertical lines indicate
the predicted 3-$\sigma$ uncertainty limits for the event.
The second (brown) curve plotted is an estimate of the relative transparency.
An electronic copy of the data in this figure is provided.
}
\end{center}

\needspace{3\baselineskip}
\subsection{Fixed stations}

\subsubsection{SOAR}\label{sec-soarjul17}

We took data at SOAR using the Raptor Merlin camera described
in Section~\ref{sec-soarjul10}.  The occultation star was the brightest star
within the field of view because of the smaller field of view of the Raptor camera.
As a consequence, we kept the exposure time at 500 ms to accommodate
the dim comparison stars available.  Observations spanned between
2017-07-17 02:48:35 and 2017-07-17 04:48:34 UTC ,  and the signal-to-noise ratio near the middle of the observation window was 65.
We did not see any signatures due to \musn\ in the data.

\subsection{Results}

For this event, we obtained five positive occultation detections roughly in the
center of the deployed stations.  Table~\ref{tbl-jul17ev} lists the measured timings
of these events.  Figure~\ref{fig-0717ch} shows a plot of the combined geometry
between the stations and the occultation timings.  These observations clearly
showed that \musn\ was more complicated than a simple ellipsoidal
object.  Our first interpretation of these data was that \musn\ was a contact
binary shape.  In the months following the initial data reduction, we began
tracking two additional scenarios to explain the outline from this occultation.
One extra option was that the occultation happened during a mutual event between
two closely orbiting bodies and just look like a contact binary due to projection
effects.  This scenario received a lot of attention due to its implication for
the New Horizons flyby: the spacecraft pointing was effectively set to look at the center of
mass.  With a binary, that point would be in between the two objects and the New Horizons spacecraft
might see nothing.  The last option was simply a very irregular shape.  This option
had no special implications for the New Horizons encounter and received no special
attention.  Still, the kink in the shape inferred from the T13 and T16 chords
would require a degree of non-sphericity not seen in any other objects in this
size class and was thus considered to be unlikely all along.

\begin{deluxetable}{cccccl}
\tablecaption{Occultation timings 2017 Jul 17\label{tbl-jul17ev}}
\tablewidth{0pt}
\tablehead{
\colhead{Team ID}&
\colhead{UT Disappearance}&
\colhead{UT Reappearance}&
\colhead{Length [km]}&
\colhead{Offset [km]}
}
\startdata
T17& 03:50:06.234& 03:50:06.551&   7.6&   24.1\\
T13& 03:50:05.690& 03:50:06.785&  26.4&   12.2\\
T16& 03:50:05.488& 03:50:06.283&  19.2&    8.9\\
T05& 03:50:04.782& 03:50:05.530&  18.0&    3.9\\
T02& 03:50:06.037& 03:50:06.412&   9.0& $-$0.2\\
\enddata
\tablecomments{\scriptsize
All times are on 2017 July 17.  Offset is relative to the last
pre-event prediction.}
\end{deluxetable}

\begin{center}
\includegraphics[scale=0.55]{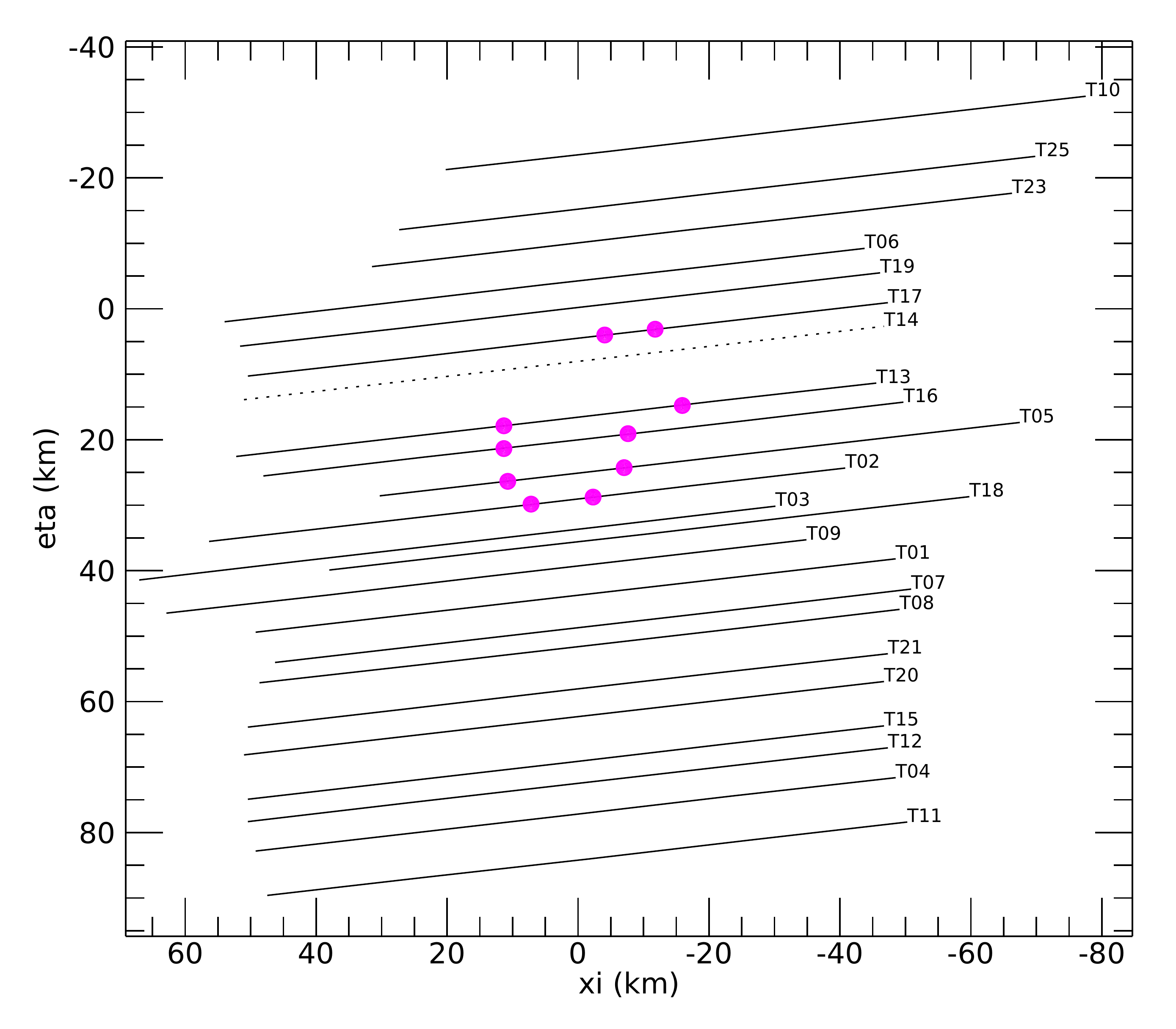}
\figcaption{\label{fig-0717ch}
Geometry plot of the 2017-07-17 occultation.
Each line shows a track of the topocentric position of \musn\ relative
to the star for 30 seconds around the time of the event and is labeled with
the team number.  The stagger
between curves is due to the variation on longitude of the observing sites.
The magenta dots indicate the interval where we observed the occultations. 
The one dotted curve did not yield timing information because of
mis-pointing at the time of the event.
The team numbers
may be cross-referenced with Table~\ref{tbl-jul17sta}.
}
\end{center}

\section{2018 August 4 Event}

Shortly following the success of the 2017 occultations, we searched the Gaia
catalog \citep{gai16} for additional occultation opportunities in 2018.  We
identified an event on 2018-08-04 involving a star of magnitude G=13.38.
Figure~\ref{fig-0804globe} shows the final predicted ground-track for this event.
We chose to deploy our telescopes to Colombia and S\'en\'egal. S\'en\'egal
would be the solitary choice for deployment most times of year since its
weather is generally much drier and more free of clouds.  However, this event
occurred during their annual rainy season.  Climatic considerations
indicated a roughly 50\% chance of clear skies in either location.
However, given the location of the ground track, we expected teams to be much
more mobile in S\'en\'egal due to simpler terrain.
Thus we sent twenty-one of the total twenty-four systems to
S\'en\'egal for the main deployment effort.  We sent the other three systems
to Colombia along with a few extra QHY cameras to be used on local telescopes. 

\begin{center}
\includegraphics[scale=1.0]{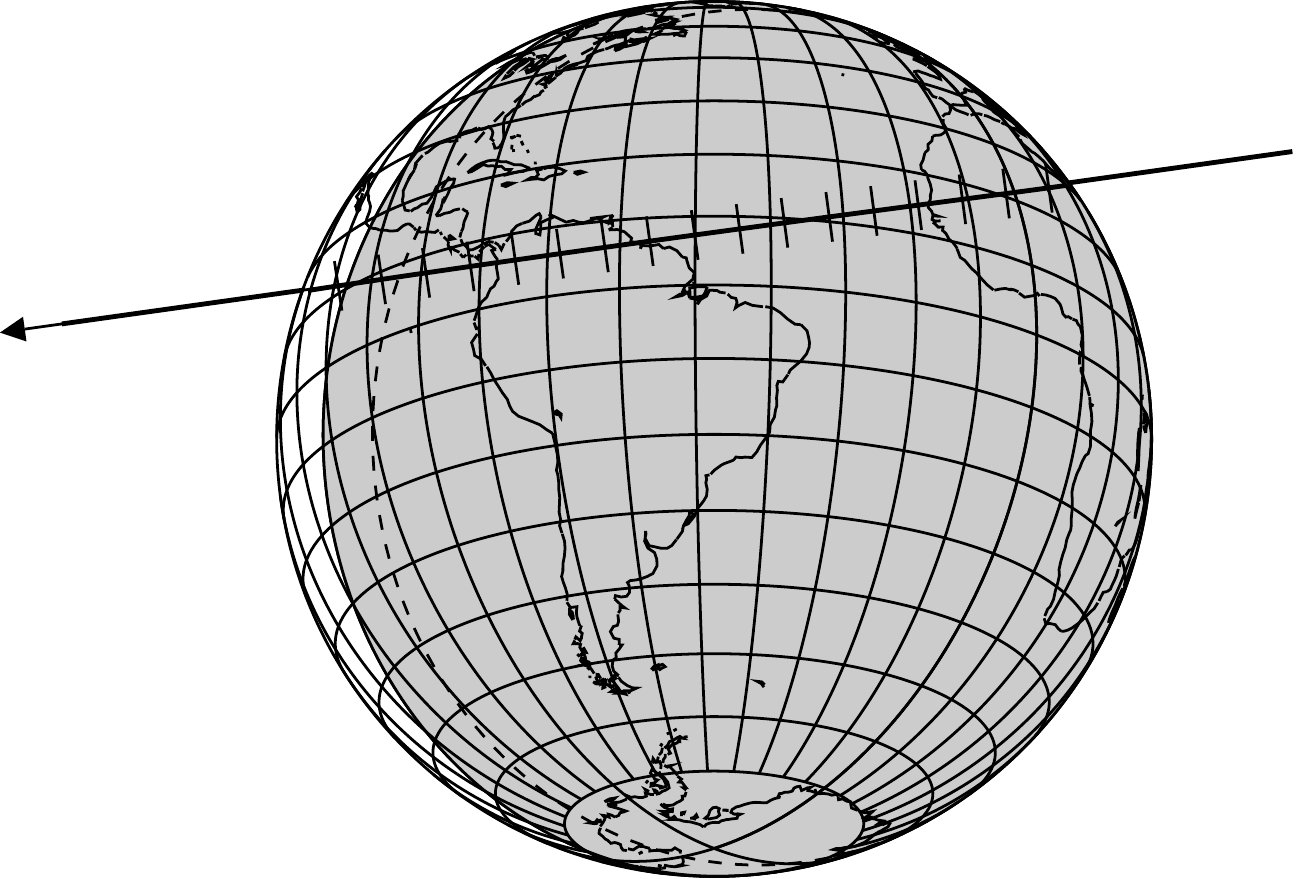}
\figcaption{\label{fig-0804globe}
Global view of 2018-08-04 occultation prediction.
The figure shows the Earth as seen from \musn\ at the time of
geocentric closest approach.  The Sun is below the horizon in the
regions shaded gray.
The dashed line shows where Sun is at -12$^\circ$ altitude.  The solid line indicates the predicted ground-track with
the width drawn to scale for a 30 km diameter object.  The arrowhead indicates the direction of motion and the ticks are spaced at
30 second intervals from 01:20:00 to 01:28:00 UT.  Shadow velocity was
21.4 km s$^{-1}$.  A 58\% illuminated Moon was 108\mydeg\ away from the target
at the time of the event.
}
\end{center}

\subsection{Prediction}

The prediction for this event was potentially better than for any of the 2017
occultations.  The observational arc was extended by additional HST data; further,
the successful occultation itself added a new astrometric constraint that was
more than an order of magnitude more constraining than any single HST observation.
The formal uncertainty for the occultation prediction was 13~km cross-track and
1.8~sec (36 km) down-track.  This uncertainty was a useful guide
provided our assumption of
a single body was true.  The size shown by the 2017 data was roughly 30$\times$14 km.
We had no basis to predict what the projected shape would be in 2018 or what the
orientation would be.  The prior result could only then suggest the minimum
(14 km) or maximum (30 km) cross-track extent.  In the case of the binary
scenario, the chance of observing during a mutual event a second time was low
and we had to be concerned with the implications of there being two separated
bodies to cover.  With two bodies, we expected diameters between 14 and 20 km
requiring tighter coverage.  However, there could also be a significant
offset between the two bodies relative to the barycenter by tens of km or more.
Thus, even though we had an excellent prediction of the center of mass
relative to the target star, that knowledge was insufficient to decide on the
deployment strategy.  Instead, our decisions were driven by what we didn't know
about \musn, a new regime for stellar occultation predictions and deployments.

\subsection{Deployment}

Table~\ref{tbl-aug04sta} summarizes the final deployment locations.  The goal
for all teams was to observe from a location along an assigned line at a fixed
distance from the centerline and within 500~m of that line.  We spaced these tracks 
at 4~km intervals centered symmetrically around the predicted centerline.  This
spacing would insure two chords on a 10~km body and the number of stations covered
almost 120~km in the cross-track direction to better cover the close binary case.
The case of the contact binary would thus be covered by more than 4$\sigma$
relative to the prediction.

\begin{deluxetable}{cP{5cm}cccccl}
%\footnotesize
\tablecaption{Observing Stations and Teams for 2018 Aug 4\label{tbl-aug04sta}}
\tablewidth{0pt}
\tablehead{
\colhead{ID}&
\colhead{Team}&
\colhead{Latitude}&
\colhead{E Longitude}&
\colhead{Elevation}&
\colhead{FWHM}&
\colhead{Sky}&
\colhead{Comments}\\
& &
\colhead{(deg)}&
\colhead{(deg)}&
\colhead{(m)}&
\colhead{(pixels)}&
\colhead{(counts)}
}
%\colnumbers
\decimals
\startdata
T01& M. Buie, M. Kaire, A. Dieng&                      $+$15.621668& $-$16.246225&   50&  3.7&  1950& some data, high extinction\\
T02& D. Dunham, C. Carter, L. Sow&                     $+$15.663333& $-$16.258917&   36&  6.1&  1983& no tracking and high extinction\\
T03& J.-L. Dauvergne, R. Smith, O. Diouf&              $+$15.363200& $-$16.420182&   30& \nodata& \nodata& no data\\
T04& B. Keeney, T. Legault, O. Bathiery&               $+$15.234900& $-$16.085017&   44& \nodata& \nodata& no data\\
T05& A. Rolfsmeier, C. Ferrell, A. Traore&             $+$15.764722& $-$16.259167&   16&  3.0&  1634& high extinction\\
T06& J. Keller, T. Finley, C. Bop&                     $+$15.323632& $-$16.259835&   41&  5.8&  1861& high extinction\\
T07& F. Colas, M. Grusin, S. Mbaye&                    $+$15.710556& $-$16.268611&   50&  4.2&  1758& some data, high extinction\\
T08& W. Hanna, R. Ballet, B. Diop&                     $+$15.554953& $-$16.294078&   41&  3.5&  1751& good data\\
T09& C. Olkin, J. Jewell, S. Gueye&                    $+$15.414722& $-$16.413889&   46& \nodata& \nodata& no data\\
T10& J. Desmars, I. Smith, D. Diakhite&                $+$15.100417& $-$16.053806&   34& \nodata& \nodata& no data\\
T11& S. Porter, S. Moss, D. Ndiaye&                    $+$15.186048& $-$16.083462&   50& \nodata& \nodata& no data\\
T12& J. Regester, A. Ocampo, G. Faye, B. Yanni&        $+$15.141700& $-$16.073083&   40& \nodata& \nodata& no data\\
T13& C. Birnbaum, J. Salmon, D. Dieng&                 $+$15.013611& $-$16.012222&   40& \nodata& \nodata& no data\\
T14& P. Tamblyn, A. Resnick, I. Gueye&                 $+$15.155167& $-$16.609417&   41&  4.7&  1954& good data\\
T15& J. Turner, J.  Samaniego, L. Toure&               $+$15.818180& $-$16.245810&   20&  5.1&  2000& no tracking\\
T16& A. Verbiscer, J. Mackie, M. Faye&                 $+$15.871544& $-$16.236944&   17&  3.6&  1831& some data, high extinction\\
T17& L. Wasserman, D. Baratoux, M. Ndiaye&             $+$15.914017& $-$16.261633&   10&  3.4&  1963& some data, high extinction\\
T18& A. Zangari, J. Dunham, M. Camara&                 $+$15.086944& $-$16.665194&   44&  4.8&  1791& useable data, moderate extinction\\
T19& P. Hinton, S. Tower, G. Dorego&                   $+$15.051117& $-$16.040150&   50& \nodata& \nodata& no data\\
T20& R. Leiva, T. Blank&                               $+$06.002778& $-$74.556944&  146& \nodata& \nodata& no data\\
T21& A.~Olsen, K.~Nowicki, D.~Rojas&                             $+$05.909444& $-$74.560556&  173& \nodata& \nodata& no data\\
T22& H. Throop, K. Getrost&                            $+$06.147778& $-$74.611944&  138& \nodata& \nodata& no data\\
T23& B. Andersen, M. Mbaye, A. Ba&                     $+$15.271792& $-$16.536416&   37&  2.4&  1872& no useful data\\
T24& M. Skrutskie, P. Edwards, M. Dieng&               $+$15.491630& $-$16.331110&   45& 1.5& 2260& no data\\
X1&  J. Castro, L. Wu, M. Gaviria&                     $+$6.296111& $-$75.330000& 2190& \nodata& \nodata& no data \\
X2& J. Zuluaga, J. Galvez, M. Ruiz, A. Torres, Y. Roman, P. Cuartas, J. Suazo, L. Ocampo& $+$6.244722& $-$75.551389& 1671& \nodata& \nodata& no data\\
X3& A. Vicini, A. Molina, K. Londo\~no&                $+$6.051667& $-$73.85278& 1671& \nodata& \nodata& no data\\
X4& A. Caycedo, N. Caycedo, G. González, F. Moreno, F. Tamayo, K. Sepulveda, F. Tamayo&    $+$6.002222& $-$73.5552787& 1735& \nodata& \nodata& no data\\
X5& R. Joya, C. Triana, L. Manzano&                    $+$5.827500& $-$73.607222& 1912& \nodata& \nodata& no data\\
X6& G. Pinz\'on, H. Rojas, S. Vanegas, S. Silva, D. Rojas& $+$5.912222& $-$73.526389& 1753& \nodata& \nodata& no data\\
\nodata& E.~Torres, M.~Arango,  D.~Rond\'on Fern\'andez, M. Guarín& $+$6.267894& $-$75.566125& 1490& \nodata& \nodata& no data\\
\enddata
\tablecomments{\scriptsize
Positions are all referenced to WGS84 datum.
}
\end{deluxetable} %J. Bustos, removed from X4, Rodrigo has confirmed

\subsubsection{S\'en\'egal}

The predicted track was in the northern portion of the country.  This area was more
thinly populated than the south and also slightly further from the direction where
storms originate.  Because the choice for the base camp location was limited, we
sent six teams to the Thi\`es area and the rest deployed from Louga.  While
splitting the teams made it much harder to tightly coordinate the deployments,
our strategy to deploy to a subset of pre-planned, fixed, cross-track locations
helped.

\subsubsection{Colombia}

The track was a little north of Bogot\'a.  During the first days we collected and
checked out the equipment and held practice sessions with all local and non-local
observers.  Our mobile teams worked close to the track for dress rehearsal night
and event night to avoid excessive driving.  The desire was to have sites
interleaved with the coverage in S\'en\'egal but it was difficult to optimize track
locations for Colombia because of geographic constraints.  

\subsection{Observations}

On the night of the event in S\'en\'egal, a storm developed and moved northward in the
hours just before the observations.  The southern sites of the deployment were
either rained out or totally clouded out.  The weather to the north had local
clouds that affected telescope setup and caused strong variability in transparency.
We took all data with 0.25-second integration times.  Only two sites were unaffected
by clouds at the appulse mid-time: T08 and T14.  The data from T08 show no obvious
occultation signal.  The data from T14 show a clear dropout of the target star for
4 consecutive frames.

In Colombia, the dress-rehearsal night was successful for setting up and taking
practice data.  Had the event been this night, these teams would have observed the
event.  On the actual night of the event a large storm rolled over the deployment
area, preventing any data collection.

\subsection{Data Reductions}

We reduced the data for these observations in essentially the same way as the
2017 data (see Section~\ref{sec:2017datared}).  Figure~\ref{fig-lc0804} shows the
data from teams whose data could be processed.  T08 and T14 provided the two best
datasets and both were straightforward to process requiring no special
treatment.

The T18 data were very challenging.  All of the data suffered from extinction
due to clouds at roughly 30\% of the signal level found at the other two sites.
Also, one of the frames during the middle of the occultation happened to be at an
instant of a guiding correction or tracking glitch and the PSF looks like a
dumbbell shape.  For this case the automatic tools treated each end of the PSF
as a separate source and the resulting stacked PSF was a very poor representation
of the image.  This problem required a manual edit of the source list to remove
the second copy of each source giving a much better representation of the image
PSF.  This new PSF was fit to the sources and then a new PSF stack was generated
from the improved positions.  This second-generation PSF was then fit to the stars
and the target as usual.

The extinction suffered by the challenging datasets was non-trivial.  For instance,
the signal from the target star was usually not even visible in the images.  Normal
photometric extraction tools that depend on measuring a position and a flux could
not retrieve the occultation signal.  However, we can compute an accurate location
for the star with our astrometric knowledge of the image and the image positions
of catalog stars.  This computed position enables us to retrieve the flux with a
constrained fit that treats the location as a given.  The T18 data have a
distinctive set of consecutive images where the fitted flux was consistent with
zero, indicating an event.  The temporal coincidence between T14 and T18 gives us
high confidence that the measurements from T18 do, in fact, represent measurements
of the limb of the object.  The constraints provided by the data from T01, T07,
T16, and T17 are minimal, but also not likely to be important given the miss
recorded by T08.

\begin{center}
\includegraphics[scale=0.55]{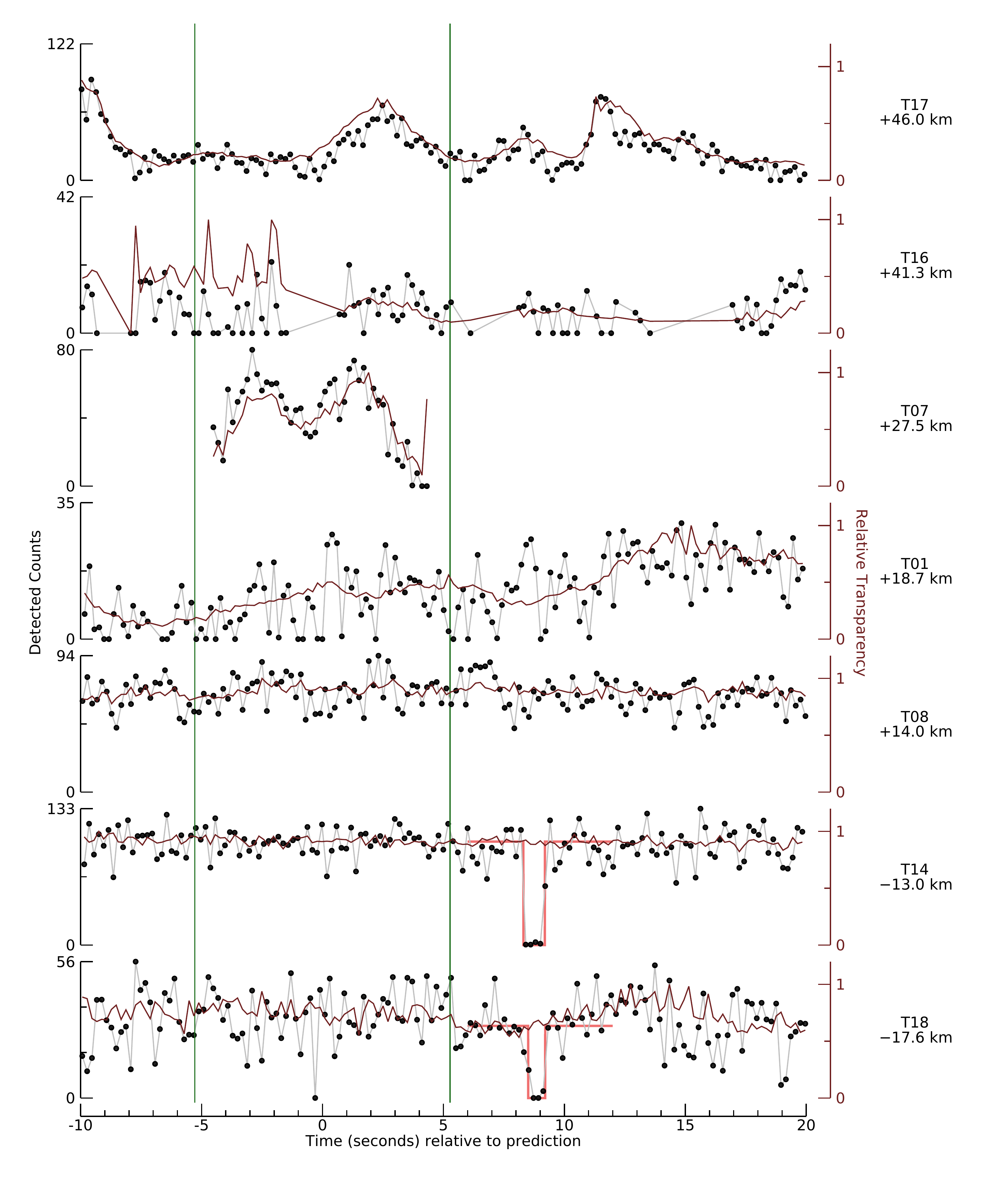}
\figcaption[figlc0804.pdf]{\label{fig-lc0804}
Observations from 2018 Aug 04 occultation.
The figure shows the lightcurves
collected by the mobile stations.  Each sub-plot is labeled
on the right with the team number and the cross-track offset.  The team numbers
may be cross-referenced with Table~\ref{tbl-aug04sta}.
The plots indicate the signal level from each station -- higher numbers
indicate higher signal levels.  The green vertical lines indicate
the predicted 3-$\sigma$ uncertainty limits for the event.
The second (brown) curve plotted is an estimate of the relative transparency.
Two of these curves show an overlain solid-body model used to extract
occultation timings.
An electronic copy of the data in this figure is provided.
}
\end{center}

\needspace{2\baselineskip}
\subsection{Results}

We obtained two positive occultation detections and one unambiguous
miss for this event.
Figure~\ref{fig-lc0804} shows the useful data from the campaign.
Table~\ref{tbl-aug04ev} lists the measured timings of the two chords.  From this
we see that the cross-track offset is comparable to the size of the object
and is consistent with the uncertainty of the prediction (13~km).
The miss was useful to rule out a second object at the location probed by T08,
14~km north of the predicted centerline.  However, this
constraint was weaker than the implication of getting the successful occultation
observations so close to the predicted position.  Additionally, both chord
lengths are longer than the minimum dimension seen in 2017.  These data had
to have hit the larger lobe or body.  With the continued improvements in the
orbit estimate of \musn\ the contact binary scenario was really considered to
be a formal, but not likely, possibility.  For the binary case, the cross-track
offset would imply that the position angle of the system was along the track and
yet we didn't see a second object.  In the end, explaining both the 2017 and 2018
data with a binary system required many special and unlikely circumstances
and thus the contact binary option was recognized as the more likely explanation.
A more quantitative analysis of the data would have been interesting but
there was no time to do so before New Horizons arrived at the target and
definitively showed the object to be a contact binary \citep{ste19}.

\begin{deluxetable}{cccccl}
\tablecaption{Occultation timings 2017 Aug 04\label{tbl-aug04ev}}
\tablewidth{0pt}
\tablehead{
\colhead{Team ID}&
\colhead{UT Disappearance}&
\colhead{UT Reappearance}&
\colhead{Length [km]}&
\colhead{Offset [km]}
}
\startdata
T14& 01:21:30.206& 01:21:31.094&  19.0& $-$13.0\\
T18& 01:21:30.604& 01:21:31.308&  15.1& $-$17.6\\
\enddata
\tablecomments{\scriptsize
All times are on 2018 August 04.  Offset is relative to the last
pre-event prediction.}
\end{deluxetable}

\subsection{Astrometry\label{sec-astrometry}}

A very powerful result of any successful occultation is in constraining
the position of the object relative to the occulted star.  This constraint
is in physical (ie., km) scale units and can easily be as good as 1~km or better.
Since the scale (km/arcsec) increases linearly with distance, these positional
constraints are especially tight in angular quantities.  For instance,
at the time of the July 17 event 1~km was equivalent to 32 microarcsec.
Before the release of the Gaia DR2 catalog, this information was of little
use since the positions of the stars were not very well known.  We can see
from the values given in Table~\ref{tbl-stars} that the position of the July 17
star was good to about 5-6~km.  Even with Gaia DR2, astrometry that is derived
from a measured occultation position at this distance is limited by the catalog.

Given that our occultation-derived astrometry was limited by the catalog,
we did not put special effort into extracting an optimized position for \musn.
Prior to the New Horizons encounter,
we took our reference position from the center of observed chords
that were the closest to the center of the body.

For the July 17 observation, we used the center of the longest chord from
station T13 and the coordinate of the
occultation star (given in Table~\ref{tbl-stars}) from the topocentric location
for site T13 (given in Table~\ref{tbl-jul17sta}) at the mid-time between
disappearance and reappearance (Table~\ref{tbl-jul17ev}).  This derived time
is thus 2017-07-17 03:50:06.238 UT\null.  The result for the August 4 event
is not quite as good since we only have two chords, but even here the
uncertainty in the star position dominates.  As before, we used the position
of the star to define the location of the object as seen from the T14 site
(longest chord) at the mid-time of the event, 2018-08-04 01:21:30.650 UT (see
Table~\ref{tbl-aug04ev} for disappearance and reappearance times).

With the then assumption and now knowledge of a single body, these occultation-based
positions provided an exceptionally strong constraint on the mean motion of
\musn, or equivalently the semi-major axis of its orbit.  This result gave us
the most important piece of information needed for a successful spacecraft encounter and
that is the heliocentric distance.  This constraint also helped to better
separate out bad astrometry from the HST data set.  The consequences of this
improvement are discussed in the next section as well as final astrometry that
takes advantage of the post-encounter shape model for \musn.

\needspace{2\baselineskip}
\subsection{Combining Event Constraints\label{sec-combine}}

A key component of this effort was to provide an orbit estimate for the
navigation of New Horizons in preparation for its encounter with \musn\ on
2019 Jan 01 \citep{ste19}.  The pre-occultation predictions and the
post-observation reconstructed ground-tracks demonstrate the evolution of this
preparation.  The first four rows of Table~\ref{tbl-errors} provide a summary
that shows the final pre-event prediction uncertainties for each occultation
campaign.  In this case, the improvement is due to each updated orbit estimate.
We experienced the largest improvement between the June 2017 and July 2017 events
because the HST lightcurve campaign added a substantial amount of astrometry just
prior to the events.  Starting in July 2017, each of our successful occultation
observations, with their much higher precision positional measurements, also
improved the orbit estimate.  The final four rows of Table~\ref{tbl-errors} show
the formal uncertainties of the reconstructed post-dictions using all of the HST
and occultation data as well as the orbit estimate delivered to the New Horizons
project.

The orbit ID ``may25a'' refers to the orbit based on all HST data up through and
including data taken on 2017 May 25.  This was the final orbit solution prior to
the first observing campaign.  ``May25a'' was based on 83 good observations
providing a total astrometric arc of almost three years.  This was our first
orbit reduced against the Gaia DR2 catalog (first pre-release version).  The
orbit with ID ``lc1'' was the first to include all of the HST lightcurve
campaign data and was used to target SOFIA for its flight.  This version was
based on 187 HST data points.  Subsequent work in the next couple of days led
to improved bad-point filtering largely driven by the requirement that the
photometry associated with the astrometry must all be consistent rather than
filtering based on astrometric residuals alone.  This led to the orbit denoted
``lc1gr'' based on 197 data points which was used to target the third
occultation campaign.  The marked reduction in the prediction errors was
crucial to the success of the July 17 effort.  The orbit with ID ``ey3jul1''
included numerous improvements in data reduction and new observations.  This
orbit was based on 230 observations from HST up through 2018 July 1 and
included the 2017 July 17 occultation measurement.  The final Gaia DR2 release
\citep{gai18} further improved the orbit along with a new pipeline image
product from STScI that eliminated trails in the images that can strongly
affect measurements of all sources, especially those on the faint end of the
detectable range.  We discovered these ``flc'' image products as an
unanticipated bonus during a data-processing detour working on astrometric
measurements of 1I/`Oumuamua \citep{mic18}.  The `Oumuamua observations
exhibited faint SNR levels comparable to those seen with \musn\ but with very
different geometric circumstances.  This overlapping effort lead to a few more
improvements in astrometric processing of the HST data.  The ``ey3jul1'' orbit
has the smallest range uncertainty of those shown in Table~\ref{tbl-errors}.

The final orbit ID ``ey7'' was the last orbit provided to New Horizons for
navigation support that did not include any New Horizons data.  This orbit was
based on 195 HST observations and two occultations, MU20170717 and MU20180804.
The overall improvement for post-dictions for all events is evident.  Timing and
cross-track uncertainties are significantly smaller in all cases.  Although the
range uncertainty is larger than was computed for the MU20180804 prediction, we
consider it more reliable due to the two high-precision occultation constraints
separated by slightly more than a year.

\begin{deluxetable}{cccc cccc cccc}
\tablecaption{Pre- and post-fit event uncertainties\label{tbl-errors}}
\tablewidth{0pt}
\tablehead{
& &
\multicolumn5{c}{MU69 uncertainty only}&
\multicolumn5{c}{Total uncertainty}\\
\colhead{Event}&
\colhead{Orbit ID}&
\multicolumn2{c}{Timing}&
\multicolumn2{c}{Crosstrack}&
\colhead{Range}&
\multicolumn2{c}{Timing}&
\multicolumn2{c}{Crosstrack}&
\colhead{Range}\\
& &
\colhead{(sec)}&
\colhead{(mas)}&
\colhead{(km)}&
\colhead{(mas)}&
\colhead{(km)}&
\colhead{(sec)}&
\colhead{(mas)}&
\colhead{(km)}&
\colhead{(mas)}&
\colhead{(km)}
}
\startdata
MU20170603& may25a&  3.326& 2.171& 44.455& 1.443& 2399.21& 3.327& 2.172& 44.614& 1.449& 2399.24\\
MU20170710& lc1&     0.969& 0.778& 15.365& 0.501& 3799.72& 0.975& 0.783& 16.160& 0.527& 3799.72\\
MU20170717& lc1gr&   0.874& 0.690& 12.879& 0.420& 2175.72& 0.883& 0.697& 14.246& 0.464& 2175.74\\
MU20180804& ey3jul1& 1.748& 1.224& 12.345& 0.402& 1427.05& 1.759& 1.232& 13.032& 0.424& 1427.05\\ \hline
MU20170603& ey7&     1.315& 0.858&  4.899& 0.159& 1661.76& 1.333& 0.870&  6.567& 0.213& 1661.76\\
MU20170710& ey7&     0.251& 0.201&  4.446& 0.145& 1679.80& 0.318& 0.255&  6.509& 0.212& 1679.80\\
MU20170717& ey7&     0.221& 0.175&  4.380& 0.143& 1683.10& 0.316& 0.249&  6.716& 0.219& 1683.10\\
MU20180804& ey7&     0.196& 0.138&  4.014& 0.131& 1877.90& 0.274& 0.193&  5.772& 0.188& 1877.90\\
\enddata
\tablecomments{\scriptsize
Orbit IDs are described in the text.  Uncertainties are all 1-$\sigma$.}
\end{deluxetable}

Data from four occultation campaigns provided very powerful constraints for
both the nature of \musn\ and its orbit -- both clearly of concern for the
New Horizons encounter.  The simplest and most likely explanation for the
occultation data throughout most data analysis efforts was a contact binary.
However, mission constraints pushed the analysis even further to quantify
those scenarios that could not be firmly ruled out.  This led to a parallel
track of interpretations: \musn\ could have been either a pair of objects
orbiting a mutual center of gravity or a single object with a strange shape.
Without a mutual orbit for a putative pair we needed to pursue heliocentric
orbit estimates without constraints from the occultations.  Doing so made the
HST-only orbit estimates robust against the interpretation of the nature of
the object but at a significant increase in the uncertainty of the resultant
fit.  Had the signal level of the \musn\ images from HST been a little higher
it is unlikely these issues would have surfaced.  However, the low signal level
meant the orbit fit was uncomfortably dependent on the bad-point editing and
weighting applied during the orbit fitting process.  If \musn\ were double, we
expected to see strong signatures of barycentric offsets in the
occultation-derived positions.  We could not completely rule out a
binary with just the MU20170717 event.  The MU20180804 event indicated that
the binary case was far less likely because that event happened at the right
cross-track position for a single-object scenario.  The larger than anticipated
time shift for the MU20180804 event is entirely consistent with a scenario where
the second occultation provides a much better determination of the semi-major
axis, especially since the interpretation of the other two events, MU20170603
(miss) and MU20170710 (graze), were also still completely consistent with the
conclusions.  By leaving out the occultation constraints, the four occultation
datasets (detections \emph{and} non-detection) were much harder to explain.
In the end, the inclusion of two occultation points led to the removal of some
data that were having a systematic erroneous effect on the orbit solution.  

This analysis also helped us understand the reference systems used for
astrometry and orbit analysis.  We were concerned about unrecognized
systematics between ``ground-based'' data and spacecraft navigation (radio
tracking and optical navigation).  This concern often leads to significant
underweighting of ground-based priors, and places higher demands on the
spacecraft navigation data, leading to more conservative uncertainty estimates
for spacecraft encounters.  For New Horizons, we had very little time to observe
\musn\ with spacecraft instruments prior to encounter and consequently these
priors became unusually important.  In particular, the constraint on the
heliocentric distance of \musn was dominated by the orbit estimate based on
HST and occultation data and could not be independently derived from optical
navigation until it was too late to use the information.  One element of the
cross-comparison centered on whether the Inertial Coordinate Reference Frame
(ICRF) was the same for the astrometric support catalog used for ground-based
measurements and the ICRF used for spacecraft navigation.
The Gaia mission team's attention to this issue \citep{lin18} was very important
to how the New Horizons mission used our orbit fitting results.  Even so, we used
very conservative estimates of the time-of-flight uncertainties for the encounter
planning.  However, this process indicates that for future missions the Gaia ICRF
is close enough to that used for radio tracking to allow treating them as
indistinguishable.  While this does not lessen the need for care
with the measurements themselves, the concern for this particular
systematic error has been eliminated through use of a common reference system.  

The four occultation observations also provide important constraints on the spin
state of \musn.  While we have not completed the shape-model analysis, we can
use the preliminary analysis to demonstrate how the encounter observations can be
linked with the occultation results.  The New Horizons imaging data alone are
sufficient to fit a shape model with a spin state (pole and rotation period).
The current best estimate is a J2000 pole direction of (311\mydeg,$-$42\mydeg)
and a spin period of 15.92$\pm$0.02 hours \citep{ste19}.  We can use this model
and any period consistent with the estimate to render \musn\ on the plane of the sky
as seen from the Earth at the time of each occultation.  The occultation data
from July 17 provide an unambiguous orientation of \musn\ at that time.
Within $\pm0.06$~hours of the nominal period, there are 7 discrete periods that
match the orientation on July 17 equally well, differing by a single full
rotation between the occultation and encounter for adjacent period options.
The nominal period implies 803.3 rotations over this time thus periods that match
within $\pm$3 rotations of this case are all possible.
For these seven possibilities we can
then look to the orientation at the time of the other events.  The June 3 event
mostly rules out any option where the major axis of \musn\ is perpendicular to
the shadow track but this is a weak constraint.  Similarly, the July 10 SOFIA data
can be used to rank the likelihood of each option but again the constraint is weak.
The last event from 2018 with the two chords gives the strongest constraint.
For this example case, the only period that provides a plausible match for the two
chords from 2018 is a period of 15.9380 hours.  Assuming the rotation is located
to within 10\mydeg\ by the July 17 data, the period would then be good to roughly
0.0005 hours, an improvement by a factor of 100.  The full analysis is left for
future work that will fully integrate the encounter imaging with the occultation
profiles.

Figure~\ref{fig-evsum} shows an example of how the current shape model matches
the occultation data using this updated period.  In this figure are shown
views of \musn\ based on the shape model from \citet{ste19} and projected to
the time and location of the occultation observation.  The dates for each view
are indicated in each sub-panel.  The projected scale is given for a sky-plane
view of the object in the J2000 coordinate system.  The field of view of each
rendered view is about 2 mas.  Relevant observations for each event are shown
by the colored lines.  Non-detections do not intersect the body and are
shown in red.  The other lines (green and yellow) are for sites recording
an occultation.  The green lines are a special marker to indicate the reference
chord that was used for astrometry (described in \S\ref{sec-astrometry}).  In the case
of the 2017-07-10 data, there are two options to match the detection with the
body and shown with dashed and solid lines.  The solid line is the most likely
based on the current orbit of \musn\ and the Gaia DR2 position of the star.
Note that the star position uncertainties are all about 4 km in the frame of
reference shown in this figure.

\begin{center}
\includegraphics[scale=0.45]{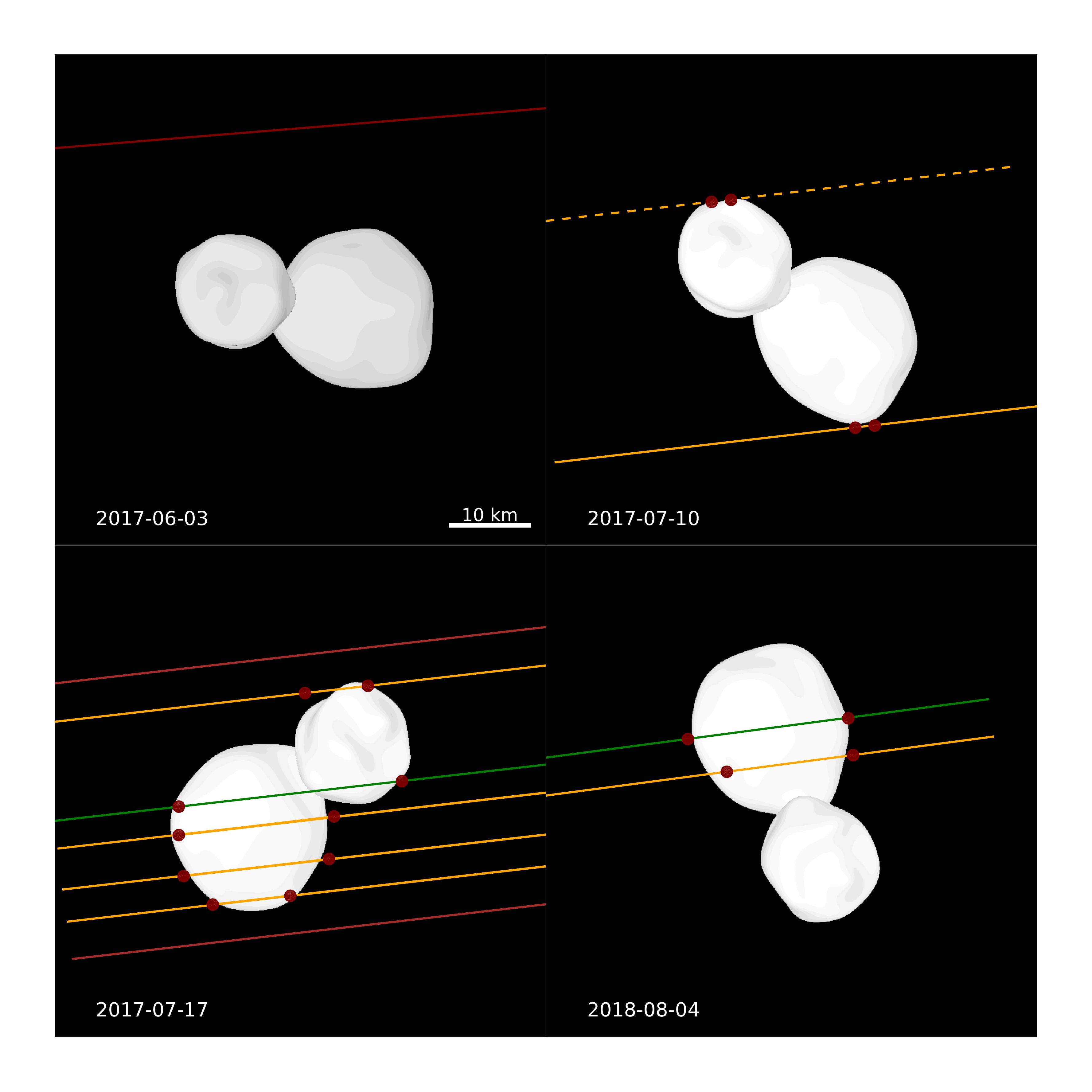}
\figcaption{\label{fig-evsum}
Comparison of shape model with occultation data.
The projected shape and orientation of \musn\ is compared to the occultation data
for all four events in a J2000 sky-plane view.  Red lines indicate observations
that did not record an occultation.  Green and yellow lines are those that did.
The red dots denote where the star either disappeared or reappeared.  The dashed yellow
line for 2017-07-10 is an alternate but less favored option.
}
\end{center}

We used the shape model as shown in Fig.~\ref{fig-evsum} to derive final astrometry
from all occultations except for the first where the object was not detected.  In
the case of the SOFIA event, we provide astrometry for both options shown in the
figure for completeness but the option where we clip the southern projected end of
the body is preferred.  The preference is driven by the better match to the shape model.
The southern end of the model (plane-of-the-sky coordinates)
comes almost to a point making it easier to get the short chord duration seen by
SOFIA.  On the northern end, there is a flat facet at that location that appears to
preclude getting such a short chord.  A more detailed shape analysis combining
spacecraft imaging data with the occultation constraints may be able to reveal the
correct interpretation.

These rendered images were converted to silhouettes, preserving the orientation
as shown in Fig.~\ref{fig-evsum}.  We do not know the mass distribution within
\musn\ but we derived an approximate location of the center of mass by finding the
area-weighted center of the silhouettes.  Any error made with this approximation is
significantly smaller than the uncertainty in the star positions.  The position of
the center is noted with respect to the center of the closest positive chord.  The
inferred astrometric positions are given in Table~\ref{tbl-ast} along with the
topocentric location for each of the measurements.

\begin{deluxetable}{ccccccc}
\tablecaption{Occultation Astrometry\label{tbl-ast}}
\tablewidth{0pt}
\tablehead{
\colhead{Site ID}&
\colhead{UTC}&
\colhead{Latitude}&
\colhead{E Longitude}&
\colhead{Altitude}&
\colhead{R.A.}&
\colhead{Dec.}\\
\colhead{}&
\colhead{}&
\colhead{[deg]}&
\colhead{[deg]}&
\colhead{[m]}&
\colhead{}&
\colhead{}
}
\startdata
SOFIA& 2017/07/10 07:49:05.50& $-$16.387247&  $+$184.961026& 12354.8& 19:00:41.620047& $-$20:38:44.53765\\
T13& 2017/07/17 03:50:06.24& $-$45.651700&  $-$67.645600&   481.0& 19:00:08.291521& $-$20:39:37.98218\\
T18& 2018/08/04 01:21:30.96&  $+$15.086944&  $-$16.665194&    44.0& 19:04:21.478510& $-$20:35:36.54727\\ \hline
SOFIA$^*$& 2017/07/10 07:49:05.50& $-$16.387247&  $+$184.961026& 12354.8& 19:00:41.620011& $-$20:38:44.53676\\
\enddata
\tablecomments{\scriptsize
R.A. and Dec. are given in EME2000 coordinates and are topocentric locations.
All topocentric locations are referenced
to the WGS84 datum. SOFIA$^*$ is the alternate, non-preferred, position for that data
set.}
\end{deluxetable}

\section{Discussion}

The results from this occultation-based effort brings this method to an
entirely new level.  Two key technological advances and one programmatic
approach were crucial to our success.  First and foremost, the Gaia astrometric
catalog precision has lived up to expectations and as a result, opened the
door to the study of the outer Solar System through occultations.  It is
now possible to predict occultations well in advance of the event to permit
large-scale targeted experiments on small and distant objects.  Based on our
\musn\ experience, occultations of transneptunian objects (TNOs) down to a diameter of 10-20 km
can be pursued with weather being the only uncontrollable risk factor
that stands in the way of success.

The second key technological advance is, at long last, the availability of
the perfect occultation camera.  The QHY174M-GPS is a perfect balance of
cost, build quality, and capability that combines rigorous timing accuracy
with a fast and sensitive detector.  The use of sCMOS technology eliminates
the need for a shutter and minimizes deadtime.  While there are even faster
sCMOS-based camera systems, this one is good enough with only $\sim$1~ms
deadtime per read.  The cost of the system is also a fundamentally valuable
advantage.  These cameras cost a total of \$26,400 to outfit all telescopes
while the previous generation
of frame transfer CCD systems could be expected to cost \$660,000 just for
the detector with extra costs for providing accurate timing.  Another element
of the systems was the design and cost of our telescopes, again an off-the-shelf
commercial product.  The 40-cm Skywatcher telescopes are easy to setup and
use under the conditions required of an occultation observation.  Our full
set of telescopes cost only \$74,800 and each is individually cheap enough that
complete replacement of a failed system is a viable option.

The capabilities of these systems, combined with their cost, allowed us to
consider an entirely new programmatic approach using a large set of mobile
equipment.  Our occultation deployment didn't break any records for the number
of involved systems or observers.  What was unusual was the size of the team
that was deployed as a single coordinated effort and all following the same
direction.  The operational methodology that emerged through our three large
mobile deployments was one of strategic placement of each and every station to
optimize the overall experiment and its chances for success.  The level of central
coordination was far stronger for our campaigns than has been typical of prior
efforts.  Our experience shows this to have been an effective strategy and one
that is especially important for mapping out the shape of small objects such as
\musn.  Objects in this size range are particularly well-suited to centrally
organized deployments.

This project provides an interesting example for the use of the Gaia star catalog by
providing a homogeneous set of data that are all referenced near catalog epoch
star positions.  The challenge of both successful occultation observations
and the New Horizons encounter was met despite the short observational arc on \musn.
The full details of the orbit and error determination are discussed in
\citet{por18}.  However, some general lessons learned from that work are worth
summarizing here. The event prediction uncertainties provided here are dependent
on capturing all sources of noise and proper handling of any priors.  The hardest
part of this process is capturing or constraining systematic errors.  We now see
that the missed occultation on our first attempt was clearly the result of
such unrecognized systematics.  Our best estimate of the source of these early
systematics is the filtering applied to reject bad astrometric measurements.
Bad point rejection became easier as our dataset increased; however, we lacked
sufficient data prior to 2017 June 3 to know for sure which measurements were bad.
To check this conjecture, we ran a test case that used the final filtering
used for ``ey7'' but restricted the orbit fit to the data available prior to
June 3.  Had that prediction been used with the exact deployment strategy
from that attempt, we would have been successful.
We suspected this was the case at the time but there was little we could do about
it.  We had insufficient ground-track coverage because of the finite number of sites
ground-track uncertainty and our size estimates at the time.  Doubling the spread,
while possible, would have resulted in spacing that was too coarse.  Given
available data at the time, it was unlikely we could have obtained a better
outcome for June 3.  Even so, the June 3 deployment provided an important
non-detection constraint as well as other logistical benefits.  This first effort
was absolutely essential as a full-up field test of this scale of deployment.
Without the June 3 event we would have severely compromised the chances of success
for the July 17 campaign.

Our experiences highlight a particularly useful occultation observing strategy.
In many cases, particularly with smaller targets, the first occultation is the
hardest to get.  Trying to get full size and shape information on this first
event makes the task even harder by requiring many more stations.  Instead
of trying to get it all on the first attempt, we can chose to cover an event
with relatively sparse coverage with the expected goal being to get just a single
chord.  This single chord can provide a much higher precision astrometric
measurement than can be achieved with the usual direct imaging observations.
For instance, one ground-based image can be expected to collect a position
good to 50~mas.  A successful occultation, even a single-chord dataset, can
easily get a position good to a factor of 100 better at 0.5~mas.  With simple
$\sqrt{N}$ scaling, getting a factor of 100 improvement to match would require
10,000 such images.  Most of these small TNOs can only be seen with large
(4-10~m) telescopes.  Sufficient access to these large apertures is not at all likely and this effort would serve for just
one object.  A more efficient way to proceed is to break down the occultation
work into two components: one to get the first chord with a low-cost deployment
strategy (eg., fixed-site network, possibly robotic) and then follow up with a high-density
mobile campaign.  It is also important to keep in mind that the high-density
campaigns are expensive -- comparable to the cost of five nights on one of
the Keck
Telescopes in Hawaii.  Thus it is well worth every practical effort to make
the most of what could be rare major coordinated campaigns unless
there is a change in the way large occultation efforts can be funded.  So far,
this type of deployment has only been possible when it is important for
supporting a spacecraft mission.

\section{Conclusions}

This work represents an unprecedented level of effort across many agencies,
projects, and international borders.  We carried out a very large centrally
organized occultation deployment effort which resulted in an unprecedented
investigation of a small body in the outer solar system.  The combination of a
new generation of occultation cameras with integrated timing, together with
multiple, large, mobile telescopes, supported by the new Gaia star catalog, has
demonstrated we now have the means to investigate this class of objects more
deeply with high-spatial density chords.

Our occultation results clearly showed that \musn\ was a contact binary.  A far
deeper perspective is added when combined with the results from the New Horizons
flyby \citep{ste19}.  With this new-found understanding, future occultations can
help answer the question of how unique or typical \musn\ is, as well as
investigating other dynamical classes of objects in the Kuiper Belt.  Surveys
that work in the thermal infrared can provide some of this context.  Only
occultations can probe hundreds or thousands of such bodies and provide
statistics on the fraction of objects that are tight binaries and contact
binaries.

On a practical note, the first occultation of this size of body is always likely
to be difficult.  Mobile ground deployments are very powerful and the only way to
get high-density multi-chord observations of a specific object, but they are also
an expensive undertaking.  Any other means by which even a single-chord observation
can be collected will lead to a substantial improvement in predictions of
subsequent events, and serve to make large ground deployments more effective.

This work demonstrates that a new pathway for understanding the Kuiper Belt
is now opened.  The combination of large mobile occultation deployments,
astrometry from large telescopes or HST, and the amazing Gaia star catalog
all combine to enable these investigations, and will be limited only by
desire and funding.

\begin{acknowledgements}

This project would have been impossible without support from many
groups and individuals.  We had exceptional support from the US State
Department (John Fazio, US Embassy in Argentina; James Garry, Heath Bailey, Cheikh Oumar Dia, US Embassy in S\'en\'egal),
CONAE (Felix Menicocci, Stan Makarchuk), and SOFIA (William Reach, Anil Dosaj,
Paul Newton, James Less, Stephen Koertge, Kimberly Ennico, Karina Leppik),
SOFIA/FPI$+$ (Enrico Pf\"uller, J\"urgen Wolf, and Manuel Wiedemann).
Special thanks go to Tony Barry for rapid turn-around updates
to his SEXTA reader for timing verification, QHY for updated camera firmware,
and Robin Glover for updates to the SharpCap software.
The discovery and subsequent astrometric images were made possible by
the Hubble Space Telescope and the wonderful staff at the Space Telescope
Science Institute (STScI).  Extracting high-precision astrometry from the HST data
was made possible by early release of Gaia DR2 data.
This work has made use of data from the European Space Agency (ESA) mission
{\it Gaia} (\url{https://www.cosmos.esa.int/gaia}), processed by the {\it Gaia}
Data Processing and Analysis Consortium (DPAC,
\url{https://www.cosmos.esa.int/web/gaia/dpac/consortium}). Funding for the DPAC
has been provided by national institutions, in particular the institutions
participating in the {\it Gaia} Multilateral Agreement.

The local logistical support from Argentina, especially from the municipality
of Comodoro Rivadavia, was critical to the success of the observations
on 2017 July 17.  Support from the Government of S\'en\'egal and the Ministry
of Higher Education, Research and Innovation (MESRI) was critical to the success
of the observation on 2018 Aug 4.  

This paper uses observations made at the South African Astronomical Observatory (SAAO).

Josselin Desmars acknowledges the funding from the European Research Council
under the European Community's H2020 (2014-2020\slash ERC Grant Agreement No. 669416 ``LUCKY STAR'').

The French participants were supported by the French Space Agency (CNES), the 2014-2020\slash ERC Grant Agreement No. 669416 ``LUCKY STAR'' and the French National Research Institute for Sustainable Development.

Salma Sylla acknowledges support from the Africa Initiative for Planetary and Space Science (\url{http://africapss.org}) and the Uranoscope de France (\url{https://uranoscope.org}).

Pablo Santos-Sanz acknowledges financial support by the European Union's Horizon
2020 Research and Innovation Programme, under Grant Agreement No. 687378, as part
of the project ``Small Bodies Near and Far'' (SBNAF) and also acknowledges financial
support from the State Agency for Research of the Spanish MCIU through the ``Center of
Excellence Severo Ochoa'' award for the Instituto de Astrofisica de Andalucia
(SEV-2017-0709).

We would like to thank the Asociaci\'on de Astronom\'ia de Colombia.
In South Africa, we would like to thank the many kind people who allowed strangers
to set up telescopes on their property.  Amanda Zangari and Christian Carter would
like to thank the kind strangers who helped out when their truck was stuck on the
side of the road.

This paper is based in part on observations obtained at the Southern Astrophysical Research
(SOAR) telescope, which is a joint project of the Minist\'{e}rio da
Ci\^{e}ncia, Tecnologia, Inova\c{c}\~{a}os e Comunica\c{c}\~{a}oes
(MCTIC) do Brasil, the U.S. National Optical Astronomy Observatory (NOAO),
the University of North Carolina at Chapel Hill (UNC), and Michigan State University
(MSU).  At SOAR, data acquisition was performed with a Raptor camera (visitor instrument)
funded by the Observatorio Nacional/MCTIC.

%SOFIA offical ack (https://www.sofia.usra.edu/science/publications/information-authors)
Based in part on observations made with the NASA/DLR Stratospheric Observatory
for Infrared Astronomy (SOFIA). SOFIA is jointly operated by the Universities
Space Research Association, Inc. (USRA), under NASA contract NNA17BF53C, and
the Deutsches SOFIA Institut (DSI) under DLR contract 50 OK 0901 to the University
of Stuttgart. Financial support for the SOFIA observations was provided by NASA
through award NAS297001 issued by USRA.

%GEMINI
Based on observations obtained at the Gemini Observatory, which is operated by
the Association of Universities for Research in Astronomy, Inc., under a
cooperative agreement with the NSF on behalf of the Gemini partnership: the
National Science Foundation (United States), National Research Council (Canada),
CONICYT (Chile), Ministerio de Ciencia, Tecnolog\'{i}a e Innovaci\'{o}n Productiva
(Argentina), Minist\'{e}rio da Ci\^{e}ncia, Tecnologia e Inova\c{c}\~{a}o (Brazil),
and Korea Astronomy and Space Science Institute (Republic of Korea).
Access was generously provided by Markus Kissler-Patig through Director’s Discretionary time at the Gemini Observatory,

This paper utilizes public domain data obtained by the MACHO Project, jointly funded
by the US Department of Energy through the University of California, Lawrence Livermore
National Laboratory under contract No. W-7405-Eng-48, by the National Science Foundation
through the Center for Particle Astrophysics of the University of California under
cooperative agreement AST-8809616, and by the Mount Stromlo and Siding Spring Observatory,
part of the Australian National University.

Thanks to all our other essential contributors for the
observing campaigns, both portable and fixed sites: 
Ousmane Bathiery,
Cyril Birnbaum,
Moussa Camara,
J.I.B. Camargo,
Andres E. Chapman,
Jhon Estiwar Rodr\'iguez Correa,
Charles Danforth,
B. Dickason,
Christopher Erickson,
John V. Fazio,
Lopez Larroude Luis Fernando.,
Clyde Foster,
Deiby Alejandro Garcia,
Marino Hernando Guar\'in Sep\'ulveda,
Souleymane Gueye,
Ibrahima Gueye,
Mat\'ias Samuel Herrera,
Nuria Lorena Iva\~nez,
Jack Lee Jewell,
Alejandro Kukenshener,
Thierry Legault,
Cynthia Soledad Leon,
Muzhou Lu,
Marcelo Jos\'e M\'arquez,
Alberto Orlando Martinez,
Andr\'es Molina,
John G. Moore,
Keith Nowicki,
Luis Ocampo,
Charles Triana Ortiz,
Blake M. Pantoja,
Alfonso Vicini Parra,
Mauricio Gaviria Posada,
Robert Reaves,
G. Benedetti-Rossi,
P. Saizar,
Vito Antonio Saraniti,
Isaac B. Smith,
Papa Lat Tabara Sow,
Enrique Torres,
Susan C. Tower,
Charlene Wiesenborn,
Lam Wu,
F. Binta H. M. Yanni, and
Leslie Young.

Overall funding for this project was from NASA's New Horizons project via contracts NASW-02008
and NAS5-97271/TaskOrder30 and special appreciation is due to Dr. James Green and
Dr. Lori Glaze at NASA Headquarters for their support.

\end{acknowledgements}

\allauthors

\end{document}